\documentclass[%
 reprint,
longbibliography,
superscriptaddress,
twocolumn,
 nofootinbib,
 floatfix,
 amsmath,
 amssymb,
 aps,
 prx,
]{revtex4-2}
\pdfoutput=1
\usepackage[utf8]{inputenc}
\usepackage{graphicx}%
\usepackage{dcolumn}%
\usepackage{bm}%
\usepackage[dvipsnames]{xcolor}
\usepackage[colorlinks=true,linkcolor=teal,citecolor=Maroon,urlcolor=Maroon]{hyperref}
\usepackage{physics}
\usepackage{siunitx}
\usepackage{float}
\usepackage{tikz}
\usetikzlibrary{quantikz2}
\usepackage{mathbbol}
\usepackage{enumitem}
\usepackage{styles}

\newcommand{\orcid}[1]{\href{https://orcid.org/#1}{\includegraphics[width=8pt]{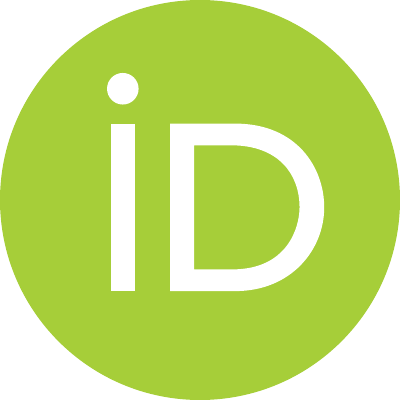}}}

\begin{document}

\title{Stroboscopic Stabilization of Cat Qubits}
\author{Timo Hillmann\orcid{0000-0002-1476-0647}
}
\email{timo.hilmann@rwth-aachen.de}
\affiliation{School of Physics, The University of Sydney, Sydney, New South Wales 2006, Australia}
\author{Franco Nori\orcid{0000-0003-3682-7432}}
\affiliation{RIKEN Center for Quantum Computing, RIKEN, Wakoshi, Saitama 351-0198, Japan}
\affiliation{Quantum Research Institute and Department of Physics, University of Michigan, Ann Arbor, Michigan 48109-1040, USA}
\author{Fernando Quijandría\orcid{0000-0002-1476-0647}}
\affiliation{RIKEN Center for Quantum Computing, RIKEN, Wakoshi, Saitama 351-0198, Japan}

\begin{abstract}
    Dissipatively stabilized cat qubits provide a promising route toward fault-tolerant quantum computation, exhibiting exponential suppression of bit-flip errors with increasing phase-space separation of the logical states, while incurring only a linear increase in phase-flip errors.
    Existing implementations rely on engineered two-photon dissipation via nonlinear coupling to a lossy environment, an approach largely confined to superconducting platforms and limited by spurious decay channels and finite dissipation rates.
    Here, we propose a fundamentally different stabilization paradigm based on repeated interactions with an auxiliary two-level system mediated by a quadratic Hamiltonian, enabling dissipative stabilization without reservoir engineering.
    Our approach overcomes key limitations of existing schemes and is compatible with a wider class of experimental platforms.
    Furthermore, it preserves the noise bias and extends to squeezed cat qubits, rendering single-photon loss errors partially correctable.
\end{abstract}

\maketitle
\date{\today}

\section{Introduction}
It is now widely accepted that quantum error correction~\cite{terhal_quantum_2015} will be necessary for nearly all practical applications of quantum computers.
However, in general, the need for quantum error correction introduces substantial overheads~\cite{beverland_cost_2021, gidney_how_2021}.
These overheads are typically significantly larger than in classical architectures, because in addition to bit-flip errors, which have a classical analog, quantum error-correcting codes have to correct phase-flip errors, which do not have a classical counterpart.
However, the ratio of bit- and phase-flip errors can be significantly altered through a careful choice of how information is encoded, potentially reducing the overheads significantly~\cite{aliferis_fault-tolerant_2008, aliferis_fault-tolerant_2009, ruiz_ldpc-cat_2025}.
For example, this can be achieved by encoding information into partially protected qubits~\cite{gyenis_moving_2021}.
Dissipatively stabilized cat qubits are a bosonic representative of a partially protected (bosonic) qubit that achieves an exponential suppression of bit-flip errors by encoding information into coherent states of the harmonic oscillator with opposite phase~\cite{mirrahimi_dynamically_2014, guillaud_repetition_2019}.
Here, we extend this work to achieve dissipative stabilization without reservoir engineering.

Bosonic qubits, in general, encode quantum information in the infinite-dimensional Hilbert space of a harmonic oscillator~\cite{joshi_quantum_2021, cai_bosonic_2021}. 
Cat qubits, in particular, are encoded into the two-dimensional subspace spanned by the coherent states $\ket{\pm \alpha}$, rendering the encoding non-local in phase space.
To protect the encoding against local perturbations such as single-photon losses, a stabilization or confinement mechanism is required.
This can be achieved either through a nonlinear Hamiltonian whose ground space, that is, the code space, is separated from the excited eigenstates by an energy gap that scales with the mean photon number $\bar{n} = \lvert \alpha \rvert^2$~\cite{puri_engineering_2017}, 
or through engineered dissipation that autonomously drives a bosonic mode into the code space~\cite{mirrahimi_dynamically_2014}. 
Here, we focus on the latter approach.
In this case, one can achieve an exponentially suppressed bit-flip rate $\Gamma_{Z} \sim \exp(-\gamma \bar{n})$, where the scaling coefficient $\gamma$ asymptotically reaches $\gamma = 2$~\cite{dubovitskii_bit-flip_2025}.
However, at the same time, the encoding becomes more susceptible to decoherence by phase-flips as $\Gamma_{X} \sim \bar{n}$.
As a result, cat qubits are an instance of a biased-noise qubit.

The performance of cat qubits can be increased by replacing the coherent states with displaced squeezed states $\ket{\pm \alpha, r}$, where $r$ is the squeezing amplitude~\cite{schlegel_quantum_2022, pan_protecting_2023}.
Intuitively, the performance gain can be understood from the exponentially reduced overlap of the displaced squeezed states compared to coherent states.
As a result, the squeezed cat qubit~\cite{schlegel_quantum_2022} has the (asymptotic) scaling coefficient $\gamma(r) = 2 e^{2r}$~\cite{hillmann_quantum_2023, xu_autonomous_2023} while the phase-flip rate remains constant for fixed $\bar{n}$.
Additionally, as the code states are not eigenstates of the bosonic annihilation operator, single-photon losses become correctable~\cite{schlegel_quantum_2022}.
Nevertheless, a stabilization protocol that is obtained by extending the two-photon dissipation protocol of cat qubits~\cite{mirrahimi_dynamically_2014} to squeezed two-photon dissipation does not protect against phase-flip errors~\cite{xu_autonomous_2023}.
In Ref.~\cite{xu_autonomous_2023}, Xu \emph{et al.} described a stabilization protocol that is capable of protecting against phase-flip errors.
However, as we detail below, the proposed physical implementations of the dissipator in Ref.~\cite{xu_autonomous_2023} have, at best, a significantly worse bit-flip rate and, at worst, an enlarged steady-state manifold that additionally contains the squeezed vacuum state.

Recently, Shitara~\emph{et al.}~\cite{shitara_exploiting_2025} proposed a protocol for stabilizing squeezed cat qubits 
by adapting a stroboscopic stabilization protocol originally developed for Gottesman--Kitaev--Preskill (GKP) qubits~\cite{gottesman_encoding_2001},
which relies on repeated interactions with an auxiliary two-level system~\cite{de_neeve_error_2022, campagne-ibarcq_quantum_2020, royer_stabilization_2020, sivak_real-time_2023, lachance-quirion_autonomous_2024, singh_towards_2025}.
However, Ref.~\cite{shitara_exploiting_2025} did not treat the stabilized qubit as a biased-noise qubit and instead analyzed the stabilization sequence in terms of the entanglement fidelity. 
Perhaps more importantly, owing to a suboptimal choice of the code's stabilizer operator, the approach in Ref.~\cite{shitara_exploiting_2025} requires relatively large photon numbers to achieve good error-correction performance. 
An additional consequence is that it inadvertently stabilizes not only the squeezed cat qubit manifold but also the squeezed vacuum state, as we explain in detail below. 

In this work, 
we therefore revisit the stroboscopic stabilization of squeezed cat qubits from the perspective of noise-biased qubits.
Notably, the stroboscopic approach preserves the qubit's noise bias while simultaneously providing partial correction of phase-flip errors. 
This offers a simple and experimentally attractive route to stabilizing squeezed cat qubits, eliminating the need for engineered two-photon dissipation and high-frequency flux modulation~\cite{lescanne_exponential_2020, rousseau_enhancing_2025, hajr_high-coherence_2024, putterman_hardware-efficient_2025}, making the protocol applicable beyond superconducting circuits, for example to trapped-ion platforms.

In particular, we discuss the role of the stabilizer operator and show that its choice has important consequences for the performance of the stabilization protocol.
Furthermore, we demonstrate that the very same stroboscopic sequences developed for GKP qubit stabilization, such as the small-big-small (sBs) sequence, 
can also stabilize ordinary (unsqueezed) cat qubits. 
To the best of our knowledge, this connection has not previously been recognized in the literature.
We perform a complete characterization of the stabilized cat qubit manifold by analyzing the influence of various typical noise processes on the performance.
Finally, we discuss the relation to previous work, such as Ref.~\cite{xu_autonomous_2023}, and make explicit how the performance of this protocol is limited.
As a side result, we also obtain accurate expressions for the bit-flip error rates of dissipatively stabilized squeezed cat qubits from first-order perturbation theory.

\section{The squeezed cat qubit}

Here we will restrict to a single bosonic mode with annihilation (creation) operators $\hat{a}$ ($\hat{a}^\dagger$). 
The position and momentum quadrature operators are defined as 
$\hat{q} = (\hat{a}^\dagger + \hat{a})/\sqrt{2}$ and $\hat{p} = i(\hat{a}^\dagger - \hat{a})/\sqrt{2}$ respectively, 
which satisfy the commutation relation $[\hat{q}, \hat{p}]=i$.

We begin by defining the displaced squeezed states
\begin{align}
    \ket{\alpha, r} = \hat{D}(\alpha) \hat{S}(r) \ket{0} ,
\end{align}
where $\hat{D}(\alpha) = \exp(\alpha \hat{a}^\dagger - \alpha^* \hat{a})$ is the displacement operator, $\hat{S}(r) = \exp[(r^* \hat{a}^2 - r \hat{a}^{\dagger 2} )/2]$
is the squeezing operator, and $\ket{0}$ is the photon vacuum state.
In the remainder of this work, we are going to restrict to real values of both the displacement amplitude $\alpha$ and the squeezing strength $r$.
Squeezed cat states are defined as coherent superpositions of displaced squeezed states with opposite displacement amplitudes
\begin{align}
    \ket{\mathcal{C}^\pm_{\alpha, r}} &= \mathcal{N}^\pm_{\alpha, r} \left( \ket{\alpha, r} \pm \ket{-\alpha, r} \right) ,
\end{align}
where $\mathcal{N}^\pm_{\alpha, r} = [2(1 \pm \exp(-2\alpha^2 \mathrm{e}^{2r}))]^{-1/2}$ is the normalization constant. 
The squeezed cat qubit logical basis is defined as
\begin{align}
    \ket{0}_{\mathcal{C}} &= \frac{1}{\sqrt{2}} \left( \ket{\mathcal{C}^+_{\alpha, r}} + \ket{\mathcal{C}^-_{\alpha, r}} \right) \approx \ket{+\alpha, r} \\
    \ket{1}_{\mathcal{C}} &= \frac{1}{\sqrt{2}} \left( \ket{\mathcal{C}^+_{\alpha, r}} - \ket{\mathcal{C}^-_{\alpha, r}} \right) \approx \ket{-\alpha, r} ,
\end{align}
where the approximate sign is a consequence of the approximate orthogonality of displaced squeezed states.
We define the code mean number of photons $\bar{n}$ as the average between the mean number of photons in the
even and odd squeezed cat states
\begin{align}
    \label{eq:mean_number_photons}
    \bar{n} &= \beta^2 \left[ \coth(2 \beta^2) \cosh(2r) - \sinh(2r) \right] + \sinh^2(r) , %
\end{align}
with $\beta = \alpha \exp(r)$. In the $\alpha \gg 1$ limit, $\coth(2 \beta^2) \to 1$ and the mean number of photons is approximated by 
$\bar{n} \approx \alpha^2 + \sinh^2(r)$.

It is instructive to analyze the squeezed cat qubit through the lens of the quantum error correction conditions~\cite{bennett_mixed-state_1996,knill_theory_1997}.
For our purposes, it is sufficient to focus on the error set $\{\hat{\mathbbm{1}}, \hat{a} \}$ corresponding to the loss of a single photon, which is often considered a dominant source of errors in optical and microwave systems.
Expressed in matrix notation, 
using the projector $\hat{P}_{\mathrm{code}}$ onto the code space,
one obtains~\cite{schlegel_quantum_2022}
\begin{align}
    \hat{P}_{\mathrm{code}} \hat{a}
    \hat{P}_{\mathrm{code}} &= \alpha e^r \ketbra{\mathcal{C}^{+}_{\alpha, r}}{\mathcal{C}^{-}_{\alpha, r}}  (\cosh (r) \nu  - \sinh(r) \nu^{-1}) \nonumber  \\&+  \alpha e^r
    \ketbra{\mathcal{C}^{-}_{\alpha, r}}{\mathcal{C}^{+}_{\alpha, r}} (\cosh (r) \nu^{-1}  - \sinh(r) \nu) \nonumber \\
    &\approx \sqrt{\bar{n} - \sinh^2r} \hat{Z} - i \alpha e^{2r} e^{-2 \alpha^2 e^{2r} }\hat{Y}, \label{eq:qec_condition}
\end{align}
where  we introduced the shorthand $\nu = \mathcal{N}_{\alpha, r}^{+} / \mathcal{N}_{\alpha, r}^{-}$ and the approximation is made in the regime where $\exp(-2 \alpha^2 e^{2r}) \ll 1$.
Importantly, the second term in  Eq.~\eqref{eq:qec_condition} shows that $Y$ errors, and thus, bit-flip errors, become super-exponentially suppressed in the squeezing parameter $r$ due to the reduced overlap of the position wavefunction peaks.
As a result, even small amounts of squeezing significantly reduce the bit-flip error rates of squeezed cat qubits compared to their unsqueezed counterparts.

Perhaps more interestingly, the first term in Eq.~\eqref{eq:qec_condition} shows that single-photon losses become increasingly correctable with increasing squeezing and do not lead to an undetectable phase-flip error, as is the case for ordinary cat qubits.
Intuitively, this can be understood by noting that displaced squeezed states are not eigenstates of the annihilation operator and thus a displaced squeezed state undergoing a photon loss is partially mapped to an error subspace. 
Ideally, an error-correcting protocol utilizes this increased protection and correctly recovers population from the error subspace to the code space.

\subsection{Standard dissipative stabilization}
A common approach to protect (squeezed) cat qubits against noise such as single-photon losses is dissipative stabilization. 
Dissipative stabilization engineers a confinement mechanism
that renders the code space an attractive manifold 
such that the system will return exponentially fast (with respect to the spectral gap of the dissipation superoperator) back to the code space as a response to any perturbation.

In particular, stabilization of the logical space is achieved by engineering a dissipative dynamics with a two-fold degenerate steady-state corresponding to the logical basis. 
By noting that squeezed displaced states are eigenstates of the squeezed annihilation operator $\hat{b} = \hat{S}(r) \hat{a} \hat{S}^\dagger(r) =
\cosh(r) \hat{a} + \sinh(r) \hat{a}^\dagger$, 
with eigenvalue $\beta = \alpha e^r$, i.e., $\hat{b}\ket{\pm\alpha, r} = \pm \beta \ket{\pm\alpha, r}$, it is straightforward to show that $(\hat{b}^2 - \beta^2)\ket{\pm\alpha,r} = 0$. 
Therefore, the dissipative dynamics governed by the Gorini-Kossakowski-Sudarshan-Lindblad (GKSL) master equation
\begin{align}\label{eq:lindblad-SC}
    \dv{}{t} \hat{\rho} = \kappa_2 \mathcal{D}[\hat{b}^2 - \beta^2]\hat{\rho},
\end{align}
evolves in the stationary state to the manifold spanned by the states $\ket{\pm \alpha, r}$, that is, to the squeezed cat qubit logical space~\cite{hillmann_quantum_2023}. 
Here, $\mathcal{D}[\hat{O}] \hat{\rho} = \hat{O} \hat{\rho} \hat{O}^\dagger - \hat{O}^\dagger\hat{O} \hat{\rho}/2 - \hat{\rho}\hat{O}^\dagger\hat{O}/2$ is the GKSL superoperator, and $\kappa_2$ is the cooling rate.

For zero squeezing ($r=0$), the above dynamics reduces to that of a two-photon driven dissipative resonator in which the stabilization of cat states, the exponential suppression of bit-flip errors, logical one-qubit gates as well as quantum error correction using a repetition code have been demonstrated in superconducting circuits experiments~\cite{leghtas_confining_2015,touzard_coherent_2018, lescanne_exponential_2020, reglade_quantum_2023, putterman_hardware-efficient_2025}.

However, this stabilization cannot take advantage of the fact that single-photon loss errors are partially correctable for $r > 0$.
To see this, let us consider the squeezing-displacement transformation $\hat{T}_{\pm} = \hat{S}^{\dagger}( r)\hat{D}^{\dagger}(\pm \alpha)$ to analyze the dissipator near its fixed points.
In the displaced squeezed frame, we obtain
\begin{align}
        \label{eq:local_two_photon_dissipator}
      \hat{T}_{\pm} (\hat{b}^{2} - \beta^2) \hat{T}_{\pm}^{\dagger} = \pm 2 \alpha \hat{\delta} + \hat{\delta}^2 \approx \pm 2 \alpha \hat{\delta},
\end{align}
where we introduced $\hat{\delta} = \hat{T}_{\pm}\hat{b}\hat{T}_{\pm}^{\dagger}$, and, in the last step, we assumed that the quantum fluctuations around the fixed points satisfy $\braket{\hat{\delta}^{\dagger}\hat{\delta}} \ll 1$.
We can interpret Eq.~\eqref{eq:local_two_photon_dissipator} as cooling to the vacuum in the local frame; however, the sign dependent whether $\hat{T}_{+}$ or $\hat{T}_{-}$ is applied.
This sign difference causes dephasing of the coherences $\ketbra{\pm \alpha, r}{\mp \alpha, r}$ of the stabilized squeezed cat qubit.

As a result, even though the squeezed cat encoding is capable of correcting phase-flip errors for $r > 0$, the dissipative stabilization described above cannot exploit this capability and dephases states while returning them from the error-subspace.

\section{Lattice view of cat qubits}\label{sect:bad-GKP}

A Gottesman-Kitaev-Preskill (GKP) qubit is a bosonic encoding utilizing translation symmetry in phase space to detect and correct errors~\cite{gottesman_encoding_2001}. 
Here, we draw inspiration from it to describe a stabilization protocol that can partially correct single-photon losses.

The square-lattice GKP code is stabilized by two commuting displacement operators $\hat{S}_q = \hat{D}(i\sqrt{2\pi})$ and $\hat{S}_p = \hat{D}(\sqrt{2\pi})$. 
The logical states can be expressed as infinite sums of position eigenstates $\ket{\mu_L} \propto \sum_{n \in \mathbb{Z}} \ket{q = (2n+\mu)\sqrt{\pi}}$, $\mu = 0,1$. 
We refer to Ref.~\cite{grimsmo_quantum_2021, brady_advances_2024} for a more thorough introduction into the topic and further note that position eigenstates can be understood as squeezed states displaced along the $q$ quadrature in the infinite squeezing limit.

\subsection{The infinite energy case}
While GKP qubits exhibit discrete translational invariance across the entire phase space, squeezed cat qubits are only translationally invariant along one quadrature (the anti-squeezed quadrature).
For real $\alpha$ and $r$, squeezed cat states become superpositions of squeezed states displaced along the $q$ quadrature.
To reveal the translational symmetry of these states, we calculate the expectation value of the displacement operator along the orthogonal quadrature ($p$), i.e., displacements of purely imaginary amplitude $i \eta$ ($\eta$ real):
\begin{align}\label{eq:displacement-expectation}
    \bra{\mathcal{C}^\pm_{\alpha, r}} \hat{D}(i\eta) \ket{\mathcal{C}^\pm_{\alpha, r}} \approx \cos(2 \alpha \eta) \exp \left[ \frac{1}{2}\text{e}^{-2r} \eta^2 \right] ,
\end{align}
where the approximate sign results from neglecting rapidly decaying exponentials in $\alpha \exp(r)$.
In the infinite squeezing limit ($r \to \infty$), the squeezed cat exhibits discrete translational invariance along the $p$ quadrature.
In this limit, the minimum periodicity along the $p$-quadrature corresponds to displacements of amplitude $\eta = \pi/2\alpha$, 
\begin{align}
    \bra{\mathcal{C}^\pm_{\alpha, \infty}} \hat{D} \left(i \eta\right) \ket{\mathcal{C}^\pm_{\alpha, \infty}} &\to -1 .
\end{align}
In other words, the infinitely squeezed cat states are invariant under the action of the operator 
\begin{align}\label{eq:infinite-energy-stabilizer}
    \hat{S} =- \hat{D}(i\pi/2\alpha) ,
\end{align}
that is, $\hat{S} \ket{\mathcal{C}^{\pm}_{\alpha, \infty}} =  \ket{\mathcal{C}^{\pm}_{\alpha, \infty}}$. Even though the squeezed cat code is not a stabilizer code, because of its similarity to the GKP code, we will loosely refer to the operator $\hat{S}$ as the stabilizer operator.
Since $\hat{S}$ acts like the identity on the code space, and the Pauli algebra must satisfy $\hat{Z}_L^2 = \hat{I}$, it should be unsurprising that by performing a displacement of amplitude $i \pi / 4 \alpha$ one implements, in the infinite squeezing limit, the Pauli-$Z$ gate as
\begin{align}
    \hat{Z}_L = -i \hat{D}(i\pi/4\alpha).
\end{align}

We note in passing that in reference \cite{shitara_exploiting_2025}, to avoid the minus sign multiplying the displacement operator in the definition of $\hat{S}$, the authors instead considered the operator $\hat{S}^2$ as the stabilizer of the infinite energy squeezed cat state. 
We will later discuss issues arising from the use of the $\hat{S}^2$ operator for the stabilization of squeezed cat states.

\subsection{The finite energy case}\label{sect:lattice-finite}
Similarly to the GKP case~\cite{royer_stabilization_2020, singh_towards_2025}, starting from an ideal state (infinite squeezing), it is possible to realize a finite-energy version of it by means of a regularization operator. 
Nevertheless, because squeezed cat states consist of a finite superposition of infinitely squeezed displaced states, rendering the squeezing finite is sufficient to ensure finite energy, without the need for an additional global Gaussian envelope, as is the case for GKP states.
It is straightforward to see that this can be achieved by using ``half'' of the GKP envelope operator $\hat{E}_\Delta = \exp(-\Delta^2 \hat{p}^2 /2)$, with $\Delta = \exp(-r)$\footnote{The Gaussian envelope operator for the GKP code is $\hat{E}_\Delta = \exp[-\Delta^2 (\hat{x}^2 + \hat{p}^2) /2]$. Its integral Kernel representation in the position representation corresponds to the Mehler kernel.}, that is,
\begin{align}
    \hat{E}_\Delta \ket{\mathcal{C}^\pm_{\alpha, \infty}} = \ket{\mathcal{C}^\pm_{\alpha, r = -\ln \Delta}} .
\end{align}
Similarly, the finite energy stabilizer is
\begin{align}
    \hat{S}_\Delta &= -\hat{E}_\Delta \hat{D}\left( \frac{i\pi}{2\alpha} \right) \hat{E}^{-1}_\Delta \\
    &= -\exp \left[ \frac{\pi}{\sqrt{2}\alpha} \left( i \hat{q} - \Delta^2 \hat{p} \right) \right] \\
    &= \exp \left[ \frac{\pi}{\sqrt{2}\alpha} \left( i \hat{q} - \Delta^2 \hat{p} \right) \pm i \pi \right] , \label{eq:finite-energy-stabilizer}
\end{align}
as $\hat{D}(i \pi /2 \alpha) = \exp (i\pi \hat{q} / \sqrt{2}\alpha)$.
Contrary to GKP states which reduce to states unrepresentative of the GKP class in the zero squeezing limit ($r=0$, or $\Delta = 1$), i.e., states with no traces of translational invariance, squeezed cat states reduce to ordinary cat states. 
Similar to their infinite energy predecessors, these are fixed-phase coherent superpositions constituting eigenstates of the parity operator. 

\section{Stroboscopic stabilization}\label{sect:autonomous}

\begin{figure*}
    \centering
    \includegraphics{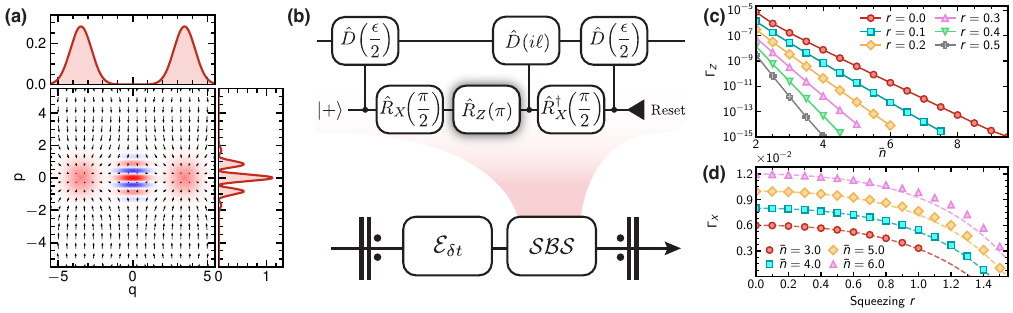}
    \caption{
    \textbf{Stabilization of cat qubits.}
    \textbf{(a)} Wigner function $W(q,p)$ and marginal distributions $\lvert \psi_q\rvert^2$ and $\lvert \psi_p\rvert^2$ of an ordinary cat state.
    \textbf{(b)} small--Big--small (sBs) stabilization sequence that is simulated interleaved with the noisy channel evolution $\mathcal{E}_{\delta t}$ for a time $\delta t$.
    The additional Pauli-$Z$ rotation of the auxiliary qubit is important to accurately represent the sign of the stabilizer~\eqref{eq:infinite-energy-stabilizer}.
    \textbf{(c)} Bit-flip rate of stabilized squeezed cat qubits. Solid lines are a guide for the eye.
    \textbf{(d)} Phase-flip rate of stabilized squeezed cat qubits.
    Dashed lines correspond to the expected phase-flip rate obtained from Eq.~\eqref{eq:qec_condition}, see \appref{app:phase_flip_rates} for details.
    In (c) and (d), the noisy channel corresponds to the single-photon loss channel with amplitude $\kappa \delta t = 10^{-3}$.
    }
    \label{fig:overview}
\end{figure*}

To engineer a dissipative process whose steady-state manifold coincides with the squeezed cat qubit code space, we can draw inspiration from the stabilization of GKP qubits.
For example, note that the operator $\hat{L}_{\Delta} = \hat{S}_{\Delta} - \hat{\mathbbm{1}}$ annihilates any state in the squeezed cat qubit manifold.
Therefore, one might expect, naively, that the dissipative dynamics governed by the Lindblad equation
\begin{align}
    \label{eq:strobo_cat_diss_goal}
    \partial_t \hat{\rho} = \Gamma \mathcal{D}[\hat{L}_{\Delta}] \hat{\rho},
\end{align}
converges towards the squeezed cat code manifold.
However, $\hat{S}_{\Delta}$ constrains the steady-state manifold only along a single quadrature axis.
Because the $q$ and $p$ quadratures are related by a Fourier transform, the resulting steady-state manifold is countably infinitely large, and solutions of dissipative dynamics are spaced by integer multiples of $2\sqrt{2}\alpha$ along the $q$ quadrature.
As a result, only if the initial state is located within a specific region in phase space, the dissipative dynamics generated by $\hat{L}_{\Delta}$ can converge towards the squeezed cat code manifold, see \sfigref{fig:overview}{a} and \secref{sec:shitara_discussion} for more details.
Apart from potential issues of physically realizing $\hat{L}_{\Delta}$, we also note that in the case of GKP qubits this approach exhibits convergence issues~\cite{sellem_exponential_2022}.

Alternatively, we can follow the approach in Ref.~\cite{royer_stabilization_2020} adapted to squeezed cat qubits. 
To this end, starting from the stabilizer relation, we can derive the dark state relations as follows,
\begin{align}
    \hat{S}_{\Delta} \ket{\psi} = \ket{\psi} \Leftrightarrow \hat{d}_{\Delta} \ket{\psi} = 0,
\end{align}

where
\begin{align}
    \label{eq:target_dissipator}
    \hat{d}_{\Delta} &= -\frac{i\alpha}{\pi \Delta} \log \hat{S}_\Delta \\
     &= \frac{1}{\sqrt{2}} \left( \frac{\hat{q}_{[2 \sqrt{2} \alpha ]}}{\Delta} + i \Delta \hat{p}  \right) \pm \frac{\alpha}{\Delta} .
\end{align}
Notice that the position quadrature operator $\hat{q}$ has been replaced by its modular counterpart $\hat{q}_{[m]} \equiv \hat{q} \mod m$ to keep track of the periodicity of the exponential.
Thus, instead of trying to engineer the dissipator $\hat{L}_{\Delta}$, we aim to engineer a dissipative evolution generated by $\hat{d}_{\Delta}$.

Generally, effective dissipative dynamics of the form $\partial_t \hat{\rho} = \Gamma \mathcal{D}[\hat{d}_\Delta] \hat{\rho}$ can be obtained from an effective Markovian oscillator-bath interaction 
\begin{align}
    \label{eq:int_hamiltonian_bath}
    \hat{H}_{\rm int}(t) = \sqrt{\Gamma} [ \hat{d}_{\Delta} \hat{w}^\dagger_{t} + \hat{d}^\dagger_{\Delta} \hat{w}_{t} ],
\end{align}
where $\hat{w}_t$ and $\hat{w}^\dagger_t$ are the bath annihilation and creation operators which fulfill $[\hat{w}_{t}, \hat{w}^\dagger_{t'}]= \delta(t-t')$ and $\ev*{\hat{w}^{\dagger}_{t} \hat{w}_{t'}} = 0$.
The interaction Hamiltonian~\eqref{eq:int_hamiltonian_bath} realizes an exchange of generalized excitations $\hat{d}_{\Delta}^{\dagger}$ and $\hat{w}_{t}^{\dagger}$ between the oscillator and the bath.
However, as $\ev*{\hat{w}^{\dagger}_{t} \hat{w}_{t'}} = 0$, the reverse process ($\hat{d}_{\Delta}^{\dagger}\hat{w}_t$) is forbidden and the effective dynamics become non-unitary, dissipatively cooling the oscillator into the $+1$ eigenstate of $\hat{S}_{\Delta}$ at a rate $\sim \Gamma$, see \appref{app:stroboscopic} for further details.
Nevertheless, the presence of the modular quadrature operator makes it non-trivial to engineer the interaction Hamiltonian in a physical system. 

Instead, following Ref.~\cite{royer_stabilization_2020}, the bath in Eq.~\eqref{eq:int_hamiltonian_bath} can be replaced by a two-level system that is reset after interacting with the oscillator via $\hat{H}_{\rm int}(t)$ for a short time $\delta t$, a common approach to represent standard models of dissipation~\cite{gross_qubit_2018, ciccarello_collision_2017}.
To this end, the time $\delta t$ must be sufficiently short such that the mean number of excitations transferred to the two-level system is much smaller than 1.
Then, the bath operator can be replaced by a two-level auxiliary system, $\hat{w}_{t} \to (\hat{\sigma}_x + i \hat{\sigma}_y) / \sqrt{2 \delta t}$, where $\hat{\sigma}_{P}, P \in \{x, y, z \}$ are the Pauli operators.
The repeatedly applied unitary operation between the qubit and the oscillator becomes
\begin{align}
    \hat{U} = \exp \left\{-i \sqrt{\frac{\Gamma \delta t}{2}} \left[ \frac{1}{\Delta} \left( \hat{q}_{[2 \sqrt{2}\alpha]} \pm \sqrt{2} \alpha \right)\hat{\sigma}_x   + \Delta \hat{p}\hat{\sigma}_y \right] \right\},
\end{align}
and the modularity of the operation can be achieved by choosing $\Gamma \delta t$ such that translating $\hat{q} \to \hat{q} + 2\sqrt{2} \alpha$ trivially affects the qubit state upon completion of the operation; see \appref{app:stroboscopic} for more details.
Once $\Gamma \delta t$ has been fixed, the unitary operation admits a straightforward physical implementation via a Trotter decomposition.
In this work, we are going to focus on the second-order approximation $\exponential({\hat{A}t + \hat{B}t}) \approx \exponential(\hat{B}t/2) \exponential(\hat{A}) \exponential(\hat{B}t/2)$, leading to a sequence of conditional displacements and a qubit rotation,
\begin{align}
    \label{eq:sBs_sequence_raw}
    \hat{U}_{\text{sBs}} &= 
    \exp \left( -i \frac{\pi \Delta^2}{4 \sqrt{2} \alpha} \hat{p} \hat{\sigma}_y \right)
    \exp \left( -i \frac{\pi}{2\sqrt{2}\alpha} \hat{q} \hat{\sigma}_x \right) \nonumber\\
    &\times
    \exp \left( -i \frac{\pi}{2} \hat{\sigma}_x \right) 
    \exp \left( -i \frac{\pi \Delta^2}{4 \sqrt{2} \alpha} \hat{p} \hat{\sigma}_y \right),
\end{align}
which, in this context, is commonly referred to as ``small-Big-small" (sBs) 
as the amplitude of the first and last displacements ($\propto \hat{p}\hat{\sigma}_y$) is smaller by a factor of $\Delta^2/2$ than the amplitude of the intermediate displacement ($\propto \hat{q}\hat{\sigma}_x$).
We emphasize that the qubit rotation $\exp ( -i \pi \hat{\sigma}_x/2 )$ is not present in the original protocol of Ref.~\cite{royer_stabilization_2020}, which is a consequence of the minus sign in the definition of the stabilizer operator~\eqref{eq:infinite-energy-stabilizer}.
In \appref{app:stroboscopic} we give a more detailed derivation of the sBs sequence,
as well as the first-order Sharpen-Trim (ST), and the second-order Big-small-Big (BsB) sequences.
We focus in the main text on the sBs sequence due to its superior performance in a wide range of relevant parameter regimes; see \appref{app:numerical_performance} for an extended discussion of the performance of various sequences.

Conditional displacement gates are realized in trapped-ion systems by the simultaneous driving of both red and blue sideband transitions induced by the phase modulation of a laser field due to the oscillatory motion of an ion~\cite{liu_hybrid_2026, matsos_robust_2024, mcgarry_programmable_2026}.
In circuit-QED, the sequence~\eqref{eq:sBs_sequence_raw} can be synthesized using conditional displacement gates $\hat{\text{CD}}(\alpha)$ which result from the dispersive interaction between a microwave resonator and a transmon qubit~\cite{eickbusch_fast_2022}, and unconditional qubit rotations $\hat{R}_P (\theta)$ as
\begin{align}
    \hat{U}_{\text{sBs}} &= 
    \hat{\text{CD}}\left(\frac{\epsilon}{2}\right) 
    \hat{R}^\dagger_x \left(\frac{\pi}{2}\right) 
    \hat{\text{CD}}(-i\ell) 
    \hat{R}_z (\pi) \nonumber\\
    &\times
    \hat{R}_x \left(\frac{\pi}{2}\right) 
    \hat{\text{CD}}\left(\frac{\epsilon}{2}\right)   ,
\end{align}
with
\begin{align}
    \hat{\text{CD}}(\alpha) &= \exp \left[ \frac{1}{2\sqrt{2}} (\alpha \hat{a}^\dagger - \alpha^* \hat{a}) \hat{\sigma}_z \right] ,\\
    \hat{R}_P (\theta) &= \exp (-i \theta \hat{\sigma}_P /2) ,
\end{align}
where $\alpha$ is the conditional displacement amplitude, $\theta$ is the qubit rotation angle around the axis set by the Pauli operator $\hat{\sigma}_P$, 
and we have defined 
\begin{align}
    \ell = \frac{\pi}{\sqrt{2} \alpha}, \quad
    \epsilon = \frac{\pi \Delta^2}{\sqrt{2} \alpha} .
\end{align}
In this case, the qubit is initialized in $\ket{+} = (\ket{g} + \ket{e})/\sqrt{2}$ state, see also \sfigref{fig:overview}{b}.

\subsection{Protection and correction mechanism}
The stroboscopic stabilization mechanism is designed such that it implements a dissipative dynamics, according to Eq.~\eqref{eq:strobo_cat_diss_goal}, so that the evolution converges towards the squeezed cat code manifold. 
It is instructive to consider how the protocol corrects errors explicitly.
To this end, note that a single-photon loss error $\hat{a}$ before the stabilization sequence $\hat{U}_{\mathrm{sBs}}$~\eqref{eq:sBs_sequence_raw} can be pulled through, to yield
\begin{align}
    \label{eq:sBs_error_push_through}
    \hat{U}_{\mathrm{sBs}}\, \hat{a} &= \hat{a}\, \hat{U}_{\mathrm{sBs}} 
    + \frac{\pi}{4 \alpha} \hat{U}_{\mathrm{sBs}}' 
    - \frac{\pi \Delta^2}{8 \alpha} \left\{ \hat{U}_{\mathrm{sBs}}, \hat{\sigma}_y \right\}, %
\end{align}
where $\{\bullet_1, \bullet_2 \}$ is the anti-commutator and $\hat{U}_{\mathrm{sBs}}^{\prime}$ corresponds to Eq.~\eqref{eq:sBs_sequence_raw} without the qubit rotation $\hat{R}_x(\pi)$.

Consider for now the case of infinite squeezing $\Delta \to 0$ such that the last term in Eq.~\eqref{eq:sBs_error_push_through} vanishes.
Note that in this limit, the sBs sequence reduces to the ``big'' displacement, that is, up to a global phase,
\begin{align}
     \hat{U}^{(\infty)}_{\text{sBs}} &\equiv 
    \sin \left( \frac{\pi}{2\sqrt{2}\alpha} \hat{q} \right) \hat{I} + i \cos \left( \frac{\pi}{2\sqrt{2}\alpha} \hat{q} \right) \hat{\sigma}_x .
\end{align}
Furthermore, in this limit, the displaced squeezed states become (generalized) position eigenstates, that is, $\ket{\alpha,r} \to \ket{q = \sqrt{2}\alpha}$.
Hence, in the limit of infinite squeezing, the action of the sBs sequence in the dual basis codewords is given by
\begin{equation}
    \hat{U}^{(\infty)}_{\text{sBs}} \ket{\mathcal{C}^\pm_{\alpha, \infty}}\ket{g} = 
    - \hat{Z}_L \ket{\mathcal{C}^\pm_{\alpha, \infty}} \ket{g},
\end{equation}
where we used that 
\begin{align}
    \sin \left( \frac{\pi}{2\sqrt{2}\alpha} \hat{q} \right) \ket{\mathcal{C}^\pm_{\alpha, \infty}} &= \hat{Z}_L \ket{\mathcal{C}^\pm_{\alpha, \infty}}, \\
    \cos \left( \frac{\pi}{2\sqrt{2}\alpha} \hat{q} \right)  \ket{\mathcal{C}^\pm_{\alpha, \infty}} &= 0 .
\end{align}
We note that the application of $\hat{Z}_L$ is inconsequential as it is state-independent and can be absorbed into the Pauli frame.

Thus, applying the stabilization sequence after a single-photon loss error has occurred, we have, using Eq.~\eqref{eq:sBs_error_push_through},
\begin{align}\label{eq:infinite-sBs-loss}
    \hat{U}^{(\infty)}_{\text{sBs}} \hat{a} \ket{\mathcal{C}^\pm_{\alpha, \infty} } \ket{g} &= 
    \hat{a}  \ket{\mathcal{C}^\mp_{\alpha, \infty} } \ket{g}
    + \frac{i\pi}{4\alpha} \ket{\mathcal{C}^\mp_{\alpha, \infty} } \ket{e} .
\end{align}
Notice that the excited branch of the auxiliary ($\ket{e}$) is associated with the successful correction of the photon-loss error.
Nevertheless, the probability amplitude associated with this branch, and therefore, the ability of the sBs sequence to correct for phase-flip errors, is inversely proportional to $\alpha$. 
Similarly, for finite squeezing $\Delta > 0$, the anti-commutator in Eq.~\eqref{eq:sBs_error_push_through} modifies the result such that in leading order in $\Delta$ we have
\begin{align}
    \label{eq:envelope-sBs-loss}
    \hat{U}_{\text{sBs}} \hat{a} \ket{\mathcal{C}^\pm_{\alpha, r} } \ket{g} \approx 
    \hat{a}  \ket{\mathcal{C}^\mp_{\alpha, r} } \ket{g}
    + \frac{i\pi}{4\alpha} (1 - \Delta^2) \ket{\mathcal{C}^\mp_{\alpha, r} } \ket{e},
\end{align}
consistent with Eq.~\eqref{eq:qec_condition} in the sense that single-photon losses are uncorrectable for the ordinary (unsqueezed) cat qubit $(\Delta = 1)$.

The probability of correction for an initially normalized squeezed cat state is  determined by the relative weight of the corrected branch in the final state~\eqref{eq:envelope-sBs-loss}
\begin{align}
    P(\ket{e}) &= \frac{ \left[ \frac{\pi (1 - \Delta^2)}{4\alpha} \right]^2 }{ \bar{n} + \left[ \frac{\pi (1 - \Delta^2)}{4\alpha} \right]^2},
\end{align}
with $\bar{n} \approx \alpha^2 + 1/(4\Delta^{2})$ the mean photon number in the limit of large squeezing, see Eq.~\eqref{eq:mean_number_photons}.
For large squeezing, the mean number of photons for both even- and odd-parity squeezed cat states is approximately the same and therefore, the above results hold regardless of the parity of the initial state. 
For fixed $\bar{n}$, increasing the squeezing enhances the ratio $(1-\Delta^2)/\alpha$, leading to a larger probability $P(\ket{e})$.
Nevertheless, smaller mean photon numbers generally yield larger values of $P(\ket{e})$.
In other words, in the large squeezing regime, the ability of the sBs sequence to correct phase-flip errors is expected to degrade with an increasing photon number.

A similar calculation to the above shows that photon-gain errors $\hat{a}^{\dagger}$ are always correctable.
In particular, one obtains
\begin{align}
    \label{eq:envelope-sBs-gain}
    \hat{U}_{\text{sBs}} \hat{a}^{\dagger} \ket{\mathcal{C}^\pm_{\alpha, r} } \ket{g} \approx 
    \hat{a}^{\dagger}  \ket{\mathcal{C}^\mp_{\alpha, r} } \ket{g}
    - \frac{i\pi}{4\alpha} (1 + \Delta^2) \ket{\mathcal{C}^\mp_{\alpha, r} } \ket{e},
\end{align}
such that the probability of correction decreases with increasing squeezing, but never vanishes.

\section{Simulations}
Here we analyze numerically the error correction capabilities of stroboscopically stabilized cat qubits by assessing the bit- and phase-flip error rates of initially prepared ideal code states.
We describe typical decoherence effects by evolution under a Markovian master equation of the general form 
\begin{align}
    \dv{}{t} \hat{\rho} = i [\hat{H}, \hat{\rho}] +\sum_i \kappa_i \hat{D}[\hat{L}_i]\hat{\rho},
\end{align}
with rates $\kappa_i$ and jump-operators $\hat{L}_i$.

\subsection{Oscillator errors}

We consider the stabilization circuit to be compiled into controlled displacements generated from the time evolution under the Hamiltonian
\begin{align}
    \label{eq:cd_hamiltonian}
    \hat{H}_{\mathrm{CD}} = \frac{g_{\mathrm{eff}}}{2} (\hat{a} e^{i \varphi} + \hat{a}^{\dagger} e^{-i\varphi}) \hat{\sigma}_z,
\end{align}
and instantaneous and perfect qubit rotations. 
Here, $g_{\mathrm{eff}} = \chi \alpha_0$ is the coupling constant that depends on the dispersive interaction coupling $\chi$ and the maximum displacement $\alpha_0$ during the gate activation, see Ref.~\cite{eickbusch_fast_2022} for more details.
For simplicity, we will assume $g_{\mathrm{eff}} /2 \pi = 1 \, \mathrm{MHz}$ in the following, compatible with previous experiments~\cite{eickbusch_fast_2022, sivak_real-time_2023}.
Alternatively, we may consider a unit system in which $g_{\mathrm{eff}} / 2 \pi \equiv 1$.
In this system of units, a conditional displacement $\hat{\mathrm{CD}}(\alpha)$ occurs in a time interval equal to
$\lvert \alpha \rvert$.
Thus, we may consider the total displacement length necessary in a stabilization sequence as the effective timescale of a single stabilization cycle. 

The total phase-space displacement necessary to complete an sBs cycle is 
\begin{align}
    \left( 1 + \Delta^2 \right) \frac{\pi}{\sqrt2 \alpha}, 
\end{align}
and is twice as large if one uses $\hat{S}^2$ instead $\hat{S}$ as done in Ref.~\cite{shitara_exploiting_2025} and reduces with increasing $\alpha$ and increasing squeezing $r$ as $\Delta = \exp(-r)$.

\subsubsection{Channel model}

We begin by analyzing the error-correcting properties of the protocol by considering the case of ideal stabilization followed by idle evolution for a time $\delta t$ during which the oscillator is subject to a loss or a dephasing channel with amplitudes $\kappa \delta t$ or $\kappa_{\phi} \delta t$, respectively, see also \sfigref{fig:overview}{b}. 
We repeat the protocol over multiple cycles until transient dynamics have died out and then obtain bit- and phase-flip rates by fitting an exponential ansatz $\ev{\sigma_Z}(t) \propto \exp(-\Gamma_{Z}t)$ and $\ev{\sigma_X}(t) \propto \exp(-\Gamma_{X}t)$, respectively, see \appref{app:observables} for further details.

We begin with the case of single-photon losses described by the jump operator $\hat{L}_{-} = \hat{a}$ and consider the case $\kappa_{-} \delta t = 10^{-3}$.
The results are shown in \sfigref{fig:overview}{c} for the bit-flip rate and in \sfigref{fig:overview}{d} for the phase-flip rate.
\sfigref{fig:overview}{c} demonstrates that in the studied regime the bit-flip error rates of the stroboscopically stabilized cat qubits are exponentially suppressed in the effective cat size.
Furthermore, \sfigref{fig:overview}{d} demonstrates that the stabilization protocol can partially correct phase-flip errors. 
This becomes particularly noticeable in the high squeezing regime.
In the regime of high squeezing and large mean photon number $\bar{n}$, we also observe deviations from the theoretical predictions (dashed lines in \sfigref{fig:overview}{d}), due to the finite correction probabilities of the stabilization sequence.

Notably, exponential suppression of the bit-flip error rate is also observed in the case of zero squeezing.
Furthermore, we observe empirically that the slopes of $\Gamma_{Z}$ approximately are described by $\lvert \beta_{\mathrm{eff}}\rvert ^2 \exp(-c \lvert \beta_{\mathrm{eff}}\rvert ^2) $ where $\beta_{\mathrm{eff}} = \mathrm{e}^{r} \sqrt{\bar{n} - \sinh^2r}$ and $c \approx 3$.
However, it is worth emphasizing that the accuracy of this heuristic is reduced for $r \gtrsim 0.5$.
We can qualitatively explain this observation through two related processes.

First, in the limit of an infinite effective size $\beta_{\mathrm{eff}}$, the cooling rate of the stabilization protocol is expected to vanish, inhibiting the stabilization protocol from working as expected.
Secondly, as a direct consequence, we find that the stabilized cat states do not accurately represent the target cat states. 
That is, when $\beta_{\mathrm{eff}}$ is large, (squeezed) cat states with a smaller effective size $\beta_{\mathrm{eff}}$ are stabilized.
We note that this effect does not fully account for the observed increase in the bit-flip rate relative to the heuristic fit.
Thus, the protocol, without further modifications, is not expected to reach arbitrarily low bit-flip error probabilities with increasing effective size. 

\begin{figure}[!tb]
    \centering
    \includegraphics{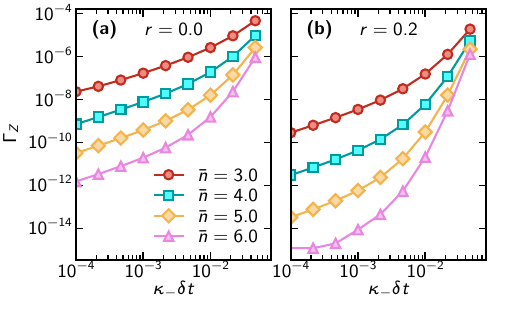}
    \caption{Bit-flip error rates of stabilized cat qubits as a function of the loss channel amplitude $\kappa_{-} \delta t$.
    The simulation assumes an ideal stabilization protocol interleaved by noisy evolution under the loss channel.
    Solid lines are a guide for the eye.}
    \label{fig:sBs_loss_sweep}
\end{figure}

Since the stabilization scheme should operate in a regime in which it can remove entropy sufficiently fast from the system, it is also relevant to explore the influence of the loss rate $\kappa_{-}$ on the protocol.
\figref{fig:sBs_loss_sweep} shows the performance of the protocol for a large range of loss amplitudes $\kappa_{-} \delta t$.
We observe that the regime in which $\Gamma_{Z} \sim \kappa_{-} \delta t$ depends on the cat state amplitude and squeezing value $r$. 
Especially for larger effective cat sizes, $\Gamma_{Z}$ shows a sharp increase with $\kappa_{-} \delta t$ due to the correction rate not being sufficiently large for the error rate induced by the loss channel.
In particular, for fixed squeezing $r$, the correction rate decreases inversely with $\bar{n}$.
In this regime, that is, the regime of large loss rates, other stabilization sequences, with increased correction rate, are potentially better suited.
However, they come with other drawbacks.
We address this specifically in \appref{app:numerical_performance}.

\begin{figure}[!bt]
    \centering
    \includegraphics{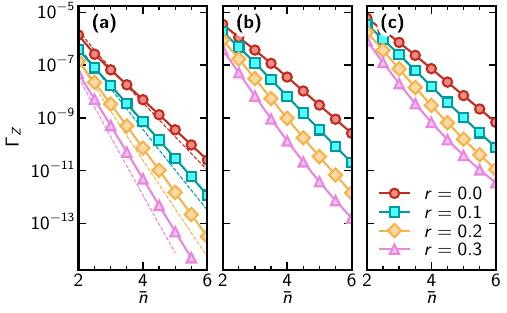}
    \caption{Bit-flip error rates of stabilized cat qubits for various values of the photon-number dephasing channel amplitude $\kappa_{\phi} \delta t$ and fixed loss amplitude $\kappa_{-} \delta t = 10^{-3}$.
    \textbf{(a)} $\kappa_{\phi} \delta t = 5\times 10^{-5}$. The dashed lines in this panel correspond to $\kappa_{\phi} \delta t = 0$ as a reference.
    \textbf{(b)} $\kappa_{\phi} \delta t = 5\times 10^{-4}$.
    \textbf{(c)} $\kappa_{\phi} \delta t = 10^{-3}$.
    } 
    
    \label{fig:sBs_timo_loss_bit_flip_dephasing}
\end{figure}

Another typical type of oscillator errors is photon-number dephasing errors, described by the Lindblad dissipator $\kappa_{\phi} \mathcal{D}[\hat{a}^{\dagger} \hat{a}]$.
Oscillator dephasing is problematic as it directly introduces leakage out of the code space.
To intuitively understand the effect of oscillator dephasing, note that for the ordinary cat qubit with $r = 0$, the jump operator $\hat{L}_{\phi} = \hat{a}^{\dagger} \hat{a}$ can be viewed as causing ``amplified'' heating of the oscillator as on the cat code subspace $\hat{L}_{\phi} \sim \alpha \hat{a}^{\dagger}$.
As a result, the effects of oscillator dephasing and oscillator heating are qualitatively similar, albeit dephasing becomes relevant already at lower rates.
Our results are summarized in \figref{fig:sBs_timo_loss_bit_flip_dephasing} which shows $\Gamma_{Z}$ for various values of $\kappa_{\phi} \delta t \in \{5 \times 10^{-5}, 5 \times 10^{-1}, 10^{-3}\}$ in the presence of simultaneous oscillator losses with amplitude $\kappa_{-} \delta t = 10^{-3}$.
Note also that dephasing errors require two stabilization cycles to be corrected, such that the regime in which $\Gamma_{Z}$ does not scale linearly with $\kappa_{\phi} \delta t$ occurs already for smaller amplitudes compared to the case of single-photon losses, see, e.g., \sfigref{fig:sBs_timo_loss_bit_flip_dephasing}{c}.

\subsubsection{Gate errors}
One could also simulate conditional displacements implemented via the Hamiltonian~\eqref{eq:cd_hamiltonian} in the presence of single-photon loss errors. 
For a duration $t_{\rm CD}$ of the conditional displacement gate, in the regime $\kappa_{-} t_{\rm CD} \ll 1$, the probability of a photon-loss event is small and can be treated perturbatively. 
Then, to first order in $\kappa_{-} t_{\rm CD}$, a photon-loss event occurring at time $\tau \in [0, t_{\rm CD}]$ during the gate splits the total displacement into two $\hat{D}(g \bar{\tau}) \hat{a} \hat{D}(g \tau)$, with  $g$ the strength of the conditional displacement gate and $\bar{\tau} = t_{\rm CD} - \tau$. 
For simplicity, we are neglecting the presence of the qubit. 
Using commutation relations, we have that
\begin{align}
\hat{D}(g \bar{\tau}) \hat{a} \hat{D}(g \tau) = \left[ \hat{a}  - g t_{\rm CD}\left(1 - \frac{\tau}{t_{\rm CD}} \right) \right] \hat{D}(g t_{\rm CD}).  
\end{align}
The first term on the right-hand side precisely corresponds to a photon loss applied after an instantaneous gate. 
The second term corresponds to a residual displacement. 
The relative strength between these two terms is roughly 
$ \sqrt{\bar{n}}/ g t_{\rm CD}$. 
For the case of the ``small" displacements in the stabilization sequence, $g t_{\rm CD} \propto {\rm e}^{-2r}/ \alpha$, and therefore the second contribution becomes negligible for finite non-zero squeezing. 
For the case of the ``big" gate, the second term yields a non-significant contribution in the $\alpha > 1$ regime explored in this work. 
Therefore, we do not expect significant deviations from the results presented here. 
Indeed, we have verified this numerically for a range of cases (not shown).
A similar analysis follows for the case of photon-number dephasing noise, see \appref{app:oscillator-dephasing}.

\subsection{Auxiliary qubit errors}
The effects of auxiliary qubit decoherence have been described in detail in Ref.~\cite{royer_stabilization_2020} for the case of GKP qubits.
It is worth quantifying the effect of decoherence channels of the auxiliary qubit system, as, for the logical $Z$ errors, the lattice constant is significantly shorter, and on the other hand, 
for the cat code, logical $X$ errors are exponentially suppressed.
As a result, it is important to establish that the bias is preserved under auxiliary dephasing mechanisms.

Hereafter we assume a model inspired by superconducting architectures, as for trapped ions, spin decoherence times are magnitudes larger than relevant gate durations.
We simulate the influence of ancilla decoherence through master equation simulations in which qubit decoherence 
can occur during the conditional displacements implemented via the Hamiltonian~\eqref{eq:cd_hamiltonian}. 
For this analysis, we assume that the auxiliary qubit rotations are perfect.
This is justified as qubit rotations can typically be implemented an order of magnitude faster than the conditional displacement gates.
In addition, we will also consider decoherence in the bosonic mode in the form of single-photon losses, which typically constitute the slowest energy-decay process in the system.

\subsubsection{Dephasing}
\begin{figure}[!t]
    \centering
    \includegraphics{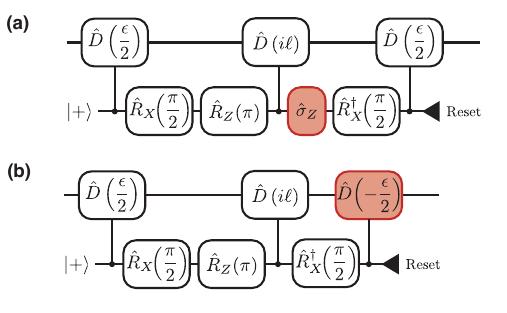}
    \caption{\textbf{(a)} A phase-flip $\hat{\sigma}_z$ error on the auxiliary qubit halfway through the sBs stabilization circuit is equivalent to \textbf{(b)} a sign-flip of the last conditional displacement.}
    \label{fig:qubit_dephasing_error_equivalance}
\end{figure}

Qubit dephasing with rate $\gamma_{\phi}$ is described by the dissipator
\begin{align}
    \frac{\gamma_{\phi}}{2} \mathcal{D}[\hat{\sigma}_z] \hat{\rho}.
\end{align}
As the controlled-displacement commutes with the generator of the dissipative evolution, it is sufficient to consider a stochastic error model in which $\hat{\sigma}_z$ is inserted before or after gates during the stabilization circuit. 
There are three non-equivalent locations. 
The first is equivalent to preparing the auxiliary qubit in the state $\ket{-}$ instead of $\ket{+}$.
It turns out that this is equivalent to performing an sBs sequence in which the sign of the ``big'' displacement is reversed.
This error is trivial as it is equivalent to the original sequence up to a translation by one lattice constant in the momentum grid.
The second location is before the reset, which does not affect the stabilized state.
The third, and perhaps the most relevant one, is an error occurring during the ``big'' conditional displacement following the auxiliary qubit rotation $\hat{R}_x(\pi/2)$, see~\figref{fig:qubit_dephasing_error_equivalance}.
The result of this error is that the sign of the last ``small'' displacement is reversed.
This causes a displacement error $\hat{D}(-\epsilon)$ along the $\hat{q}$ quadrature, which vanishes in the limit of a large cat $\alpha \gg 1$ or in the limit of large squeezing $\Delta \ll 1$.
Thus, if dephasing errors occur at a sufficiently low rate, then auxiliary qubit dephasing errors preserve the noise bias of the stabilized cat qubit. 

\begin{figure}[!t]
    \centering
    \includegraphics{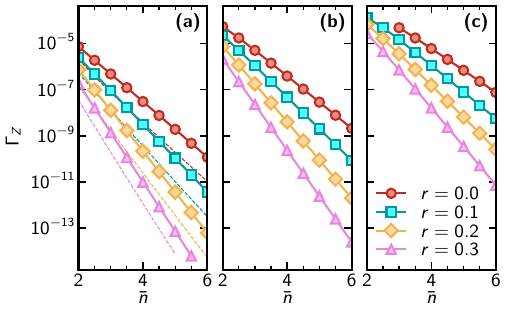}
    \caption{
    \textbf{Influence of qubit dephasing.}
    Bit-flip rate $\Gamma_{Z}$ in the presence of oscillator single photon losses at rate $\kappa_{-}$ and for different dephasing rates of the auxiliary qubit
    \textbf{(a)} $\gamma_{\phi} /\kappa_{-} = 1$,
    \textbf{(b)} $\gamma_{\phi} /\kappa_{-} = 10$,
    \textbf{(c)} $\gamma_{\phi} /\kappa_{-} = 50$.
    Dashed lines in (a) represent $\Gamma_{Z}$ in the case of an ideal auxiliary qubit $\gamma_{\phi}/ \kappa_{-} = 0$.%
    }
    \label{fig:bit_flip_qubit_dephasing}
\end{figure}

\figref{fig:bit_flip_qubit_dephasing} shows the bit-flip rate $\Gamma_{Z}$ as a function of the mean number of photons $\bar{n}$ for different values of the squeezing parameter $r$, and
for various qubit dephasing rates $\kappa_{\phi}$ at a fixed oscillator single-photon loss rate $\kappa_{-}$.
If the dephasing rate is small, that is, $\gamma_{\phi}/ \kappa_{-} = 1$, \sfigref{fig:bit_flip_qubit_dephasing}{a} shows that $\Gamma_{Z}$ is only marginally increased compared to the case of a perfect auxiliary qubit with $\gamma_{\phi} = 0$ (dashed lines). 
A larger dephasing rate $\gamma_{\phi}/ \kappa_{-} = 10$ further increases $\Gamma_{Z}$, but does so approximately linearly in $\gamma_{\phi}$ as it can be seen in \sfigref{fig:bit_flip_qubit_dephasing} {b}.
However, in \sfigref{fig:bit_flip_qubit_dephasing}{c} we can observe that 
by further increasing the dephasing rate to $\gamma_{\phi}/ \kappa_{-} = 50$, the linear behavior in $\gamma_{\phi}$ vanishes. 
In terms of our previous analysis, here the high dephasing rate leads to oscillator displacement errors that occur faster than they can be corrected by the stabilization sequence.

\subsubsection{Decay}
The case of spontaneous decay of the auxiliary qubit is distinct from the case of dephasing, as the jump operator $\hat{\sigma}_{-}= \ketbra{0}{1}$ does not commute with the controlled displacement Hamiltonian~\eqref{eq:cd_hamiltonian}.
As a result, we need to analyze the evolution
\begin{align}
    \label{eq:qubit_decay_evolution}
    \dv{}{t} \hat{\rho} = - i \comm{\hat{H}_{\rm CD}}{\hat{\rho}} + \gamma_{-} \mathcal{D}[\hat{\sigma}_{-}] \hat{\rho},
\end{align}
where $\gamma_{-} = 1/T_1$ is the decay time of the auxiliary qubit.

Before we simulate Eq.~\eqref{eq:qubit_decay_evolution} to analyze the effect of qubit decay quantitatively, we begin by considering a simplified stochastic model consisting of random bit-flips to gain intuition, following Ref.~\cite{royer_encoding_2022}.
A bit-flip on the auxiliary qubit flips the sign of the applied controlled displacement as $\hat{\sigma}_x \hat{\sigma}_z = - \hat{\sigma}_z \hat{\sigma}_x$.
Thus, if the bit-flip happens exactly at $t_{\rm CD} / 2$, the overall gate applied to the oscillator is the identity.
In general, 
if the bit-flip occurs at an arbitrary time $\tau \in [0, t_{\rm CD}]$, the resulting applied controlled displacement is given by 
$\hat{\mathrm{CD}}(\alpha_{\rm eff})$, with $\alpha_{\rm eff} \equiv \alpha (2 \tau / t_{\rm CD} - 1)$, where $\alpha$ is the desired displacement amplitude in the absence of errors.
Since the (squeezed) cat code differs from the GKP code, it is important to distinguish between bit-flips occurring during the ``small'' and the ``big'' displacement.

First, let us focus on the ``big'' conditional displacement. 
This corresponds to a stabilizer translation by $\ell$.
Bit-flip errors lead to logical $Z$ errors when they produce conditional displacements of small effective amplitude (closer to 0 than to $\ell$). 
Taking $\alpha_{\rm eff} = \vert \ell \vert/2$ as a decision boundary, logical errors occur whenever $\alpha_{\rm eff} < \vert \ell \vert/2$, that is, for bit-flip errors occurring within the time interval $\tau \in [t_{\rm CD}/4, 3t_{\rm CD}/4]$.
However, while phase-flip errors are undesirable, cat qubits are designed to protect against bit-flip errors such that this type of error preserves the noise-bias of the bosonic encoding.

\begin{figure}[!hb]
    \centering
    \includegraphics{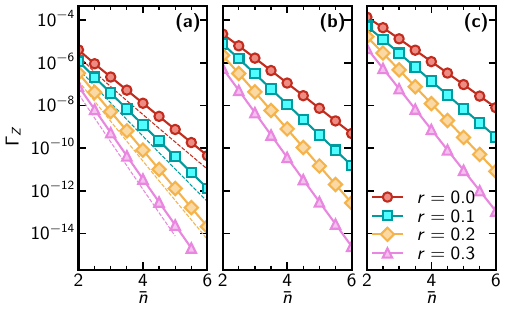}
    \caption{
    \textbf{Influence of qubit decay.}
    Bit-flip rate $\Gamma_{Z}$ in the presence of oscillator single photon losses at rate $\kappa_{-} = 10^{-3}$ and for different decay rates of the auxiliary qubit
    \textbf{(a)} $\gamma_{-} = 10^{-3}$,
    \textbf{(b)} $\gamma_{-} = 10^{-2}$,
    \textbf{(c)} $\gamma_{-} = 5 \times 10^{-2}$.
    Dashed lines in (a) represent $\Gamma_{Z}$ in the case of an ideal auxiliary qubit $\gamma_{-} = 0$ and $\kappa_{-}=10^{-3}$.
    }
    \label{fig:bit_flip_qubit_decay}
\end{figure}

Conversely, if $\alpha_{\rm eff}$ remains closer to $\vert \ell \vert$, that is, if the bit-flip error occurs outside of the time interval $[t_{\rm CD}/4 , 3t_{\rm CD}/4]$, the resulting error amounts to a small displacement around the correct stabilizer action. Such errors remain correctable as long as they occur with probability $p \approx \gamma_{-} t_{\rm CD} \ll 1$. 
On the other hand, bit-flip errors occurring during the ``small'' displacement are, in principle, more harmful as they potentially reduce the phase-space separation between logical codewords of the encoding. 
However, the displacement error that an auxiliary bit-flip error can cause is at most a fraction of the total (small) displacement amplitude $\epsilon/2 = \Delta^2 \pi/ 4\alpha$ and therefore remains correctable as long as $\Delta^2 \pi/ 4 < \alpha^2$ and bit-flip errors occur with sufficiently low probability, see also the discussion for auxiliary dephasing errors.
The simulation results shown in \figref{fig:bit_flip_qubit_decay} demonstrate that this is indeed the case.

The results shown in \figref{fig:bit_flip_qubit_decay} for auxiliary qubit decay are qualitatively similar to those in \figref{fig:bit_flip_qubit_dephasing} for auxiliary qubit dephasing.
As expected from the preceding discussion, auxiliary qubit decay leads to a smaller increase in the bit-flip rate $\Gamma_{Z}$ compared to qubit dephasing, as both decoherence channels lead to oscillator displacement errors; however, the displacement amplitude caused by auxiliary qubit decay is strictly smaller than in the case of auxiliary qubit dephasing.

\section{Discussion}

\subsection{Relation to Xu~\emph{et~al.}}

We note that Xu~\emph{et~al.}~\cite{xu_autonomous_2023} were the first to introduce an autonomous QEC protocol capable of detecting and correcting single-photon loss errors. 
Their protocol is based on the dissipative evolution
\begin{align}\label{eq:LJ-SC}
    \dv{}{t} \hat{\rho} = \kappa_2 \mathcal{D}[\hat{Z}_L (\hat{b}^2 - \beta^2) ]\hat{\rho} ,
\end{align}
which is a variation of Eq.~\eqref{eq:lindblad-SC}.
Intuitively, it should be clear that the addition of the operator $\hat{Z}_L$ fixes the issue that is pointed out around Eq.~\eqref{eq:local_two_photon_dissipator} for the scheme based on Eq.~\eqref{eq:lindblad-SC}.
Formally, this can be argued using the shifted Fock basis, see \appref{app:phase_flip_rates}.

The physical implementation, and its caveats, can be best understood in a squeezed frame $\hat{S}^\dagger(r) \hat{b} \hat{S}(r) \to \hat{a}$. 
In Ref.~\cite{xu_autonomous_2023} the authors suggested approximating the dissipator in \eqref{eq:LJ-SC} by $\hat{F}_{(c_1,c_2)} = (c_1 \hat{a} + c_2\hat{a}^\dagger )(\hat{a}^2 - \beta^2)$, that is, replacing the logical $Z$ operator 
for a linear superposition of annihilation and creation operators, where $c_1$ and $c_2$ are two real numbers satisfying $c_1 + c_2 = 1$.
Using the shifted Fock basis, it can be shown that the operator $c_1 \hat{a} + c_2\hat{a}^\dagger$ indeed yields the desired dynamics, see \appref{app:phase_flip_rates}.
Furthermore, according to the authors, the modified dissipator should not only correct for phase-flip errors but should also preserve the exponential suppression of bit-flip errors achieved by the model in Eq.~\eqref{eq:lindblad-SC}. 
However, as an extremal counter-example consider the dissipator with $(c_1,c_2) =(1,0)$, that is, $\hat{F}_{(1,0)} = \hat{a}(\hat{a}^2 - \beta^2)$.
While the kernel of the jump operator $\hat{F}' = \hat{a}^2 - \beta^2$ has rank two and is spanned by the coherent states $\ket{\pm \beta}$, the kernel of $\hat{F}_{(1,0)}$ has rank three.
That is, $\ker[\hat{F}_{(1,0)}] = \mathrm{span}(\{\ket{0}, \ket{\pm \beta} \})$. 
For $r=0$,  it is straightforward to show that in the presence of single-photon losses, the unique steady state corresponds to the photon vacuum state, which does not belong to the logical space. 
In the presence of finite squeezing, one can show numerically that the steady state also lies outside of the logical space. 
On the other hand, for the value $(c_1,c_2) =(0,1)$, the resulting operator $\hat{F}_{(0,1)} = \hat{a}^\dagger(\hat{a}^2 - \beta^2)$ does not stabilize states outside of the logical space. 

\begin{figure}[!ht]
    \centering
    \includegraphics{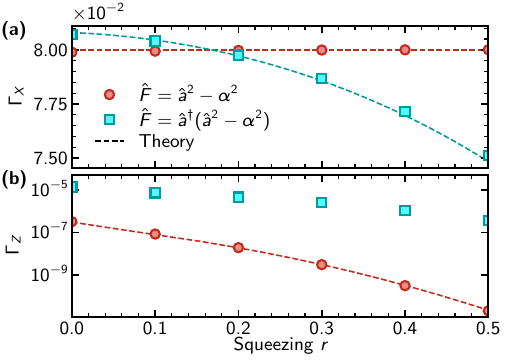}
    \caption{
    Dissipative stabilization of the squeezed cat code using different jump operators $\hat{F}$ with corresponding 
    decay rate $\kappa_2$, in the presence of single-photon losses and photon dephasing with rates $\kappa_-$ and $\kappa_\phi$, respectively. 
    \textbf{(a)} Phase-flip and \textbf{(b)} bit-flip rates of the dissipators $\hat{F}_{(0,1)}$ and $\hat{F}'$ and their theoretical predictions as dashed lines, see \appref{app:phase_flip_rates} for details.
    Here, $\kappa_{-} / \kappa_2 = 10^{-2}$, $\kappa_{\phi} / \kappa_2 = 10^{-4}$, and $\bar{n} = 4.0$.}
    \label{fig:liang_comparison}
\end{figure}

In \figref{fig:liang_comparison} we show numerical results comparing the phase- and bit-flip error rates for $\hat{F}_{(0,1)}$ and $\hat{F}'$ under simultaneous single-photon losses and photon number-dephasing.
For $\hat{F}_{(0,1)}$ the phase-flip error rate is effectively mitigated as predicted theoretically; however, the bit-flip rate is severely compromised.  %
It is not our goal to provide a detailed analysis of the dynamics generated by the dissipator $\hat{F}_{(0,1)}$, however, a simple semi-classical analysis already 
shows some differences between the dynamics governed by $\hat{F}_{(0,1)}$ and $\hat{F}'$.
For instance, the dynamics led by $\hat{F}_{(0,1)}$ possesses an additional stable point at the origin of phase space, whereas the origin is a saddle point for $\hat{F}'$. 
Therefore, a naive approximation of $\hat{Z}_L \sim c_1 \hat{a} + c_2\hat{a}^\dagger$ does not realize the desired dynamics.
The authors of Ref.~\cite{xu_autonomous_2023} did not numerically observe these issues, as all their simulations have been performed with the idealized dissipator $\hat{Z}_L (\hat{a}^{2} - \alpha^2)$ in the shifted Fock basis.

\subsection{Relation to Shitara \emph{et al.} \label{sec:shitara_discussion}}
In Ref.~\cite{shitara_exploiting_2025}, the authors studied the stroboscopic stabilization of squeezed cat states for quantum error correction, i.e., for the correction of logical phase-flip errors. 
To this end, they considered a sequence of gates (Sharpen and Trim) derived following the procedure summarized in \secref{sect:autonomous}, 
but using $\hat{S}^2$ as the stabilizer operator of the code. 
This choice was purposely made to avoid dealing with the minus sign appearing in the definition of the operator $\hat{S}$ in Eq.~\eqref{eq:infinite-energy-stabilizer}.
The main consequence of this approach is that due to the squaring, the periodicity (or modularity if we consider the logarithm of the stabilizer operators) of the stabilizer operator $\hat{S}^2$ in the position quadrature is reduced by a factor of 2 as compared to the original stabilizer $\hat{S}$. 
This has several consequences, and they can be easily visualized using phase space portraits representing the variation of the amplitude of an initially displaced squeezed state $\ket{\beta, r}$ upon the action of a single step of a stabilization sequence. 

For the sake of this discussion, we are going to restrict ourselves to the sBs sequence stabilizing the squeezed cat code defined by the parameters $\alpha$ and $r$
(in \appref{app:phase-portraits} we discuss the phase-space portraits corresponding to the ST and BsB sequences). 
Therefore, the phase space portrait corresponds to the real and imaginary parts of $\Tr(\hat{a}\, \mathcal{SBS}[\ketbra{\beta, r}{\beta, r}]) - \Tr(\hat{a} \ketbra{\beta, r}{\beta, r})$, where $\mathcal{SBS}[\hat{\rho}]$ represents the action of the sBs sequence on the bosonic state $\hat{\rho}$ after the auxiliary qubit is traced out. 

\begin{figure}[!th]
    \includegraphics{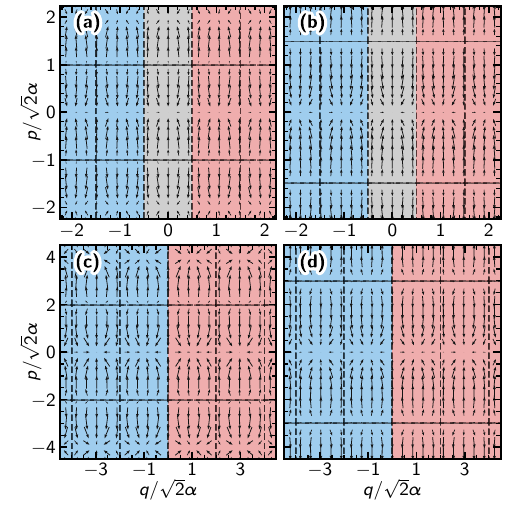}
    \caption{Phase-space portraits of Shitara \emph{et al.} protocol and ours.
    In all panels $\alpha = \sqrt{3}$, with $r = 0$ in (a) and (c), and $r = 0.2$ in (b) and (d).
    The top panels (a) and (b) correspond to the protocol with $\mu = 1$ and the bottom panels (c) and (d) correspond to the protocol with $\mu= 0$.
    Dashed lines indicate lattice and superlattice periodicity.
    The color of each cell encodes whether starting in this cell yields dynamics that yield $\ket{0_L}$ (red) or $\ket{1_L}$ upon measurement.
    The gray cells in (a) and (b) indicate attracting regions corresponding to a random logical measurement outcome in the steady state.    
    }
    \label{fig:phase_potrait}
\end{figure}

In \figref{fig:phase_potrait} we show the phase space portraits corresponding to the sBs sequence resulting from the use of the stabilizer $\hat{S}^2$ \sfigref{fig:phase_potrait}{a} and \sfigref{fig:phase_potrait}{b}, and $\hat{S}$ in \sfigref{fig:phase_potrait}{c} and \sfigref{fig:phase_potrait}{d}.
First of all, we observe that multiple attractors appear along the $q$ quadrature.
This is a consequence of the periodicity of the stabilizers. 
Let us denote $m_\mu = (1+\mu)\sqrt{2}\alpha$ the period of the stabilizer operator $\hat{S}^{(2-\mu)}$ with $\mu = 0,1$, i.e. $m_1 = 2 m_0$.
Starting from $q = \pm \sqrt{2} \alpha$, the attractors are located at $q^*_{k,+} = \sqrt{2} \alpha + k m_\mu$, $k \in \mathbb{Z}^+$ (red boxes), and $q^*_{k,-} = -\sqrt{2} \alpha - k m_\mu$, $k \in \mathbb{Z}^+$ (blue boxes), and each of the repeating cells correspond to $q \in [ q^*_{k, \pm} - m_\mu/2, q^*_{k, \pm} + m_\mu/2 ]$. 
Because the distance between the center of the logical states is equal to $m_1 = 2 m_0$, for the stabilization with $\hat{S}^2$ there needs to exist an attractor located at $q_0^*=0$, similar to the protocol 
in Ref.~\cite{xu_autonomous_2023} discussed above.
This is represented by the gray boxes in \sfigref{fig:phase_potrait}{a} and \sfigref{fig:phase_potrait}{b}. 
The presence of this attractor explains the observation in Ref.~\cite{shitara_exploiting_2025} that starting from the vacuum state, it is not possible to stabilize a squeezed cat state using the sequence derived from $\hat{S}^2$. 
This is not the case for the sequence derived from $\hat{S}$ due to its larger period. 
In fact, we have numerically verified for this sequence that, starting from the photon vacuum, it is possible to reach the code space (not shown).

A second observation is that the ``small" displacements
$\exp(-i \epsilon_\mu \hat{p} \hat{\sigma}_y /4)$, with $\epsilon_\mu = 2\pi \Delta^2/ m_\mu$ [see~\appref{app:stroboscopic}] have a period of 
$4 \pi / \epsilon_\mu = 2\sqrt{2}(1+\mu) \alpha {\text{e}}^{2r}$ along the momentum quadrature due to the presence of the qubit operator. 
Therefore, the phase portraits are also periodic along the $p$ quadrature.
Whereas the period in $p$ can be increased by increasing the squeezing $r$ (compare left and right columns in\figref{fig:phase_potrait}), 
for fixed values of $\alpha$ and $r$, 
the periods in $q$ and $p$ for the stabilization sequence resulting from $\hat{S}$ are twice those resulting from $\hat{S}^2$.
This means that the sBs sequence resulting from $\hat{S}$ has a larger domain of attraction and, furthermore, because it does not have an attractor at $q=0$, it can stabilize states with fewer photons as compared to the sequence derived from $\hat{S}^2$, see for example \sfigref{fig:overview}{a} which corresponds to a cat state with $r=0$, and $\bar{n} = 6$.

\begin{figure}
    \centering
    \includegraphics{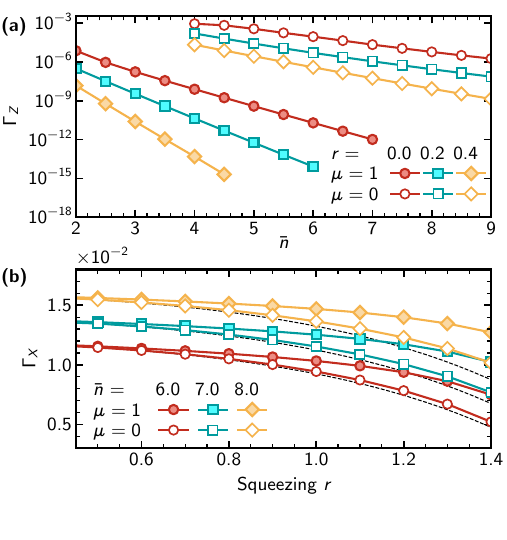}
    \caption{
    Bit- and phase-flip error rates of stabilization sequences with different lattice length.
    Simulations are performed under the assumptions of perfect and instantaneous gates interleaved by evolution under the single-photon loss channel with 
    $\kappa \delta t= 10^{-3}$.
    In, \textbf{(b)} the black dashed lines correspond to the optimally achievable phase-flip rates for a given average photon number $\bar{n}$, see \secref{app:phase_flip_rates}.
    In \textbf{(a)}, note that the protocol with $\mu = 0$ has a significantly reduced slope due to the additional solution at the origin.
    }
    \label{fig:sBs_comparison}
\end{figure}

In \figref{fig:sBs_comparison} we show the bit- and phase-flip rates for stabilized squeezed cat qubits using sBs sequences derived from the stabilizer operators $\hat{S}$ and $\hat{S}^2$.
All the above-mentioned factors contribute to significantly higher bit-flip rates for stabilized squeezed cat qubits 
using the protocol of Ref.~\cite{shitara_exploiting_2025}; see \sfigref{fig:sBs_comparison}{a}.
The different slopes of $\Gamma_{Z}$ between both protocols are worth emphasizing, as for the protocol of Ref.~\cite{shitara_exploiting_2025} they are indicative of the additionally stabilized squeezed vacuum.
While this protocol shows a slightly reduced phase-flip rate $\Gamma_X$, see \sfigref{fig:sBs_comparison}{b}, this is expected and a direct consequence of the shorter periodicity, and thus, increased cooling rate.

Finally, because the amplitudes of the ``small" and ``big" displacements are inversely proportional to $m_\mu$, the stabilization with $\hat{S}$ leads to smaller displacement amplitudes. This becomes relevant when considering physical realizations
in which conditional displacements are implemented using entangling gates of finite duration. 
Smaller amplitudes translate into shorter operation times which reduce the occurrence of errors during the gate implementation.

\subsection{Fault-tolerant architectures}
Our stabilization sequence requires the use of an auxiliary qubit system to engineer dissipative stabilization in contrast to previous methods which utilize a nonlinear coupling to a lossy bath.
As a result, one may integrate the auxiliary qubit system directly into a fault-tolerant architecture based on squeezed cat qubits.
In Ref.~\cite{putterman_hardware-efficient_2025}, the authors use (dissipatively) stabilized cat qubits concatenated with a repetition code.
However, contrary to previous proposals, e.g., Ref.~\cite{guillaud_repetition_2019}, stabilizers of the repetition code are measured via ordinary two-level auxiliary systems instead of (stabilized) cat qubits. 
Only in the latter case can syndrome extraction be performed in a bias-preserving manner~\cite{guillaud_repetition_2019}.
However, since the proposed stabilization scheme in this work does not yield an arbitrary suppression of bit-flip errors, concatenation with a (true) quantum error-correcting code is required to achieve fault-tolerance.
A proposal for measuring $Z$-type and $X$-type stabilizers via auxiliary qubit systems is naturally done in the architecture described in Ref.~\cite{putterman_hardware-efficient_2025}.
The analysis of achievable logical error rates through concatenation in this way is beyond the scope of this work.

Furthermore, we note that fault-tolerant measurements and other operations on stabilized cat qubits, such as a bias-preserving $\hat{X}_L$, can be implemented during the stabilization sequence.
Implementations of these gates carry over naturally from previous work on stabilized GKP qubits~\cite{singh_towards_2025} through the lattice view of cat qubits described in \secref{sect:bad-GKP}.

\subsection{Trapped Ions}
In trapped-ion systems, a laser field can be used to couple the ion's internal (spin) and motional (bosonic) degrees of freedom. Assuming driving fields resonant with the red and blue motional sidebands, and applying the rotating wave approximation, the resulting $n$-th order sideband interaction can be written as
\begin{align}
    \hat{H} &= g_n \hat{\sigma}_\beta \left( \hat{a}^n + \hat{a}^{\dagger n} \right) ,
\end{align}
where $\hat{\sigma}_\beta$ denotes a linear combination of the Pauli matrices~\cite{haroche_exploring_2006, wineland_experimental_1998}. 
Here, the coupling strength $g_n$ is proportional to the $n$-th power of the Lamb--Dicke parameter $\eta$, i.e., $g_n \sim \eta^n$. 
The Lamb--Dicke parameter $\eta$ corresponds to the ratio of the spatial extent of the ion's ground-state wavefunction to the wavelength of the driving field, and it is typically much smaller than unity for tightly confined ions. 

Realizing standard dissipative stabilization of cat qubits requires a nonlinear interaction of the form $\hat{a}^2 \hat{w}^{\dagger} + \hat{a}^{\dagger 2} \hat{w}^{\dagger} + \mathrm{H.c.}$ between the bosonic mode and an auxiliary lossy system to engineer the effective evolution~\eqref{eq:lindblad-SC}.
Implementing this approach with trapped ions is challenging because the spin degree of freedom is typically long-lived, while the required second-order sideband interaction is suppressed by the second power of the Lamb--Dicke parameter. 
As a result, the achievable effective two-phonon dissipation rates $\kappa_2$ are expected to be small.

Using the stroboscopic stabilization scheme proposed here, only a quadratic interaction, i.e., a first-order sideband interaction, is necessary such that the rate of stabilization is increased.
For example, for realistic parameters~\cite{mcgarry_programmable_2026}, the unitary part of the stabilization sequence can be implemented within $40$ to $120 \, \mu\mathrm{s}$, depending on the cat state squeezing and amplitude.
However, we note that in a trapped-ion system, the dominant decoherence mechanisms of the bosonic mode are motional heating and motional dephasing, both arising from electric-field noise. For instance, in a system dominated by motional dephasing~\cite{mcgarry_programmable_2026}, the cat qubit encoding might not be necessarily optimal, and encodings tailored to the physical noise mechanism, e.g., rotationally symmetric codes~\cite{grimsmo_quantum_2020, hillmann_performance_2022}, such as binomial codes and their generalizations~\cite{michael_new_2016, zeng_quantum_2026, bashmakova_bosonic_2025, gutman_squeezed-vacuum_2026},
or numerically optimized encodings~\cite{leviant_quantum_2022, zeng_approximate_2023, wang_bosonic_2026},
are likely more appropriate.
A full analysis is beyond the scope of this work.

\section{Conclusions}
In conclusion, we have performed a complete characterization of a (dissipative) stabilization protocol for cat qubits.
Our protocol is inspired by the lattice viewpoint of squeezed cat qubits through which we applied techniques originally designed for the stabilization of finite-energy Gottesman--Kitaev--Preskill qubits~\cite{royer_stabilization_2020}.
Our characterization is complete in the sense that it takes into account the biased-noise nature of cat qubits and 
incorporates decoherence mechanisms of both the oscillator and the auxiliary qubit system.
We have compared our protocol against previous proposals designed to protect against phase-flip errors and showed that both achieve this only through a significant increase in the bit-flip rate.
In this way, our proposal presents the first stabilization protocol that utilizes the properties of the squeezed cat qubit encoding almost optimally.
While our characterization is complete, it is not completely quantitative.

For future work, it is an interesting and relevant question whether it is possible to derive the expected bit-flip rates in the stabilized cat qubit manifold analytically. 
Currently, our heuristic expression for the bit-flip rates becomes inaccurate in the limit of large squeezing.
Furthermore, it is interesting to consider the question of designing better stabilization sequences with respect to observed bit-flip rates and increased cooling rates.
Initial efforts to achieve this by going to a higher-order Suzuki-Trotter approximation (fourth-order) of the total unitary evolution were unsuccessful in the simulated parameter regime.
Lastly, as mentioned in the discussion, it would be interesting to analyze the stabilization protocol in a concatenated architecture, for example based on surface codes~\cite{chamberland_building_2022} or, more generally, quantum low-density parity-check codes~\cite{berent_analog_2024}.

\acknowledgments
T.~H. acknowledges support from Defence Science and Technologies Group (DSTG) and Advanced Strategic Capabilities Accelerator (ASCA) through its Emerging and Disruptive Technologies (EDT) Program, as well as the hospitality of beta coffee, where part of this work was completed.
F.N. is supported in part by the Japan Science and Technology Agency (JST) [via the CREST Quantum Frontiers program Grant No. JPMJCR24I2, the Quantum Leap Flagship Program (Q-LEAP), the Moonshot R\&D Grant No. JPMJMS256E, and the ASPIRE program (Grant No. JPMJAP2513)], and the Office of Naval Research (ONR) Global (via Grant No. N62909-23-1-2074).
    
\appendix

\section{Bit-flip error rates of squeezed cat qubits}
Predicting the bit-flip error rates of ordinary (dissipatively stabilized) cat qubits requires
second-order perturbative calculations with respect to the nonlinear dissipative Lindbladian.
The reason is that first-order calculations for single-photon losses predict an exponential decay
$\Gamma_{   \mathrm{bit\text{-}flip}} \sim |\alpha|^2 e^{-4|\alpha|^2}$, whereas asymptotically one
observes $\Gamma_{\mathrm{bit\text{-}flip}} \sim e^{-2|\alpha|^2}$.
The reduced slope of $\Gamma_{\mathrm{bit\text{-}flip}}$ is recovered only in second-order
perturbation theory~\cite{dubovitskii_bit-flip_2025} and dominates for
$|\alpha| \geq \log(\kappa_2/\kappa_{-})/2$, the crossover point at which the second-order
contribution overtakes the first-order one.
Intuitively, single-photon loss $\mathcal{D}[\hat{a}]$ causes leakage outside the computational
subspace already at second order, so non-computational states must be included.
One finds through detailed calculations~\cite{dubovitskii_bit-flip_2025}
\begin{align}
    \label{eq:bit_flip_rate_ordinary_cat}
    \Gamma_{\mathrm{bit\text{-}flip}}
        \approx \kappa_{-} |\alpha|^2 e^{-4|\alpha|^2}
                + \frac{\kappa_{-}^2}{2\kappa_2}\, e^{-2|\alpha|^2}.
\end{align}
 
For squeezed cat qubits, bit-flip error rates can be estimated accurately already from
first-order perturbation theory.
Conceptually, instead of evaluating perturbations of the cat-code manifold induced by
single-photon losses $\mathcal{D}[\hat{a}]$, one works in a squeezed frame in which the
relevant perturbation is the squeezed-photon loss $\mathcal{D}[\hat{b}]$, where
\begin{align}
    \hat{b} = \hat{S}^{\dagger}(r)\,\hat{a}\,\hat{S}(r)
            = \cosh(r)\,\hat{a} - \sinh(r)\,\hat{a}^{\dagger}.
\end{align}
Expanding the corresponding dissipator gives
\begin{align}
    \label{eq:dissipator_squeezed_losses}
    \mathcal{D}[\hat{b}]\,\hat{\rho}
        &= \cosh^2\!r\;\mathcal{D}[\hat{a}]\hat{\rho}
         + \sinh^2\!r\;\mathcal{D}[\hat{a}^{\dagger}]\hat{\rho} \nonumber\\
        &\quad
         + \sinh r\cosh r\!\left[
               \hat{a}\hat{\rho}\hat{a}
             + \hat{a}^{\dagger}\hat{\rho}\hat{a}^{\dagger}
             - \tfrac{1}{2}\bigl\{
                   \hat{a}^{\dagger 2} + \hat{a}^{2},\,\hat{\rho}
               \bigr\}
           \right],
\end{align}
where $\{\hat{A},\hat{B}\} := \hat{A}\hat{B} + \hat{B}\hat{A}$ denotes the anticommutator.
The second term in Eq.~\eqref{eq:dissipator_squeezed_losses} corresponds to photon gain
(``heating'') and, crucially, causes leakage outside the computational subspace already at
leading order.
This is the mechanism that makes a first-order treatment sufficient.
 
Accordingly, the bit-flip rate of the squeezed cat qubit due to single-photon losses
receives three contributions, which we label ``loss'', ``heating'', and ``mixed'':
\begin{align}
    \label{eq:gamma_bit_flip_first_order}
    \Gamma_{\mathrm{bit\text{-}flip}}
        = \Gamma_{\mathrm{bit\text{-}flip}}^{(\mathrm{loss})}
        + \Gamma_{\mathrm{bit\text{-}flip}}^{(\mathrm{heat})}
        + \Gamma_{\mathrm{bit\text{-}flip}}^{(\mathrm{mixed})}.
\end{align}
 
\paragraph*{Loss contribution.}
The leading loss contribution follows directly from the first term of
Eq.~\eqref{eq:dissipator_squeezed_losses}, weighted by $\cosh^2 r$:
\begin{align}
    \label{eq:gamma_loss}
    \Gamma_{\mathrm{bit\text{-}flip}}^{(\mathrm{loss})}
        = \kappa_{-}\cosh^2\!r\;|\alpha|^2\,e^{-4|\alpha|^2}.
\end{align}
 
\paragraph*{Heating contribution.}
The heating contribution has previously been calculated in Ref.~\cite{hillmann_quantum_2023}:
\begin{align}
    \label{eq:gamma_heat}
    \Gamma_{\mathrm{bit\text{-}flip}}^{(\mathrm{heat})}
        = \kappa_{-}\sinh^2\!r\;\operatorname{csch}(2|\alpha|^2),
\end{align}
where $\operatorname{csch}(x) = 1/\sinh(x)$.
For $|\alpha|\gg 1$ this decays as $e^{-2|\alpha|^2}$, recovering the expected asymptotic
behavior of the bit-flip rate.
 
\paragraph*{Mixed contribution.}
The mixed contribution is obtained by following the first-order calculations of
Ref.~\cite{dubovitskii_bit-flip_2025}:
\begin{align}
    \label{eq:gamma_mixed}
    \Gamma_{\mathrm{bit\text{-}flip}}^{(\mathrm{mixed})}
        &\stackrel{|\alpha|\gg 1}{\approx}
           8\kappa_{-}\sinh r\cosh r\;|\alpha|^2\,e^{-4|\alpha|^2}.
\end{align}
This term is exponentially suppressed relative to the heating contribution and is therefore
subdominant at large $|\alpha|$.

\begin{figure}[!htb]
    \centering
    \includegraphics{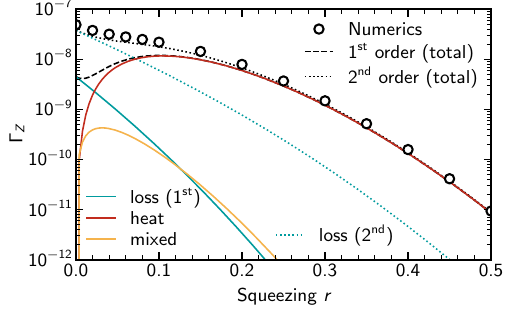}
    \caption{Bit-flip rate for the continuous dissipator $\hat{D}[\hat{b}^2 - \beta^2]$ where $\hat{b}$ is the squeezed annihilation operator.
    Markers correspond to numerically obtained bit-flip rates.
    The dashed line corresponds to the first-order contributions of Eq.~\eqref{eq:gamma_bit_flip_first_order}, with individual components shown as solid lines.
    The dotted line is identical to the dashed line but also contains second-order corrections to $\Gamma_{Z}^{\mathrm{(loss)}}$, see Eq.~\eqref{eq:bit_flip_rate_ordinary_cat}.
    Here, $\bar{n} = 4$ and $\kappa_{-} / \kappa_2 = 10^{-2}$.
    }
    \label{fig:theory_model_cont_dissipator}
\end{figure}

We demonstrate the accuracy of our model in \figref{fig:theory_model_cont_dissipator} by comparing it to numerically extracted bit-flip rates for the continuous dissipator $\kappa_2 \hat{D}[\hat{b}^2 - \beta^2]$, where $\hat{b}$ is the squeezed annihilation operator in the presence of single-photon losses at rate $\kappa_{-} / \kappa_2 = 10^{-2}$ for (squeezed) cat states with a fixed average photon number $\bar{n}= 4$.
The data shows that it is sufficient to consider first-order perturbation theory to accurately extract bit-flip rates in the regime $r \gtrsim 0.1$ in which the contribution due to $\Gamma_{Z}^{\mathrm{(heat)}}$ dominates.
We note that in the regime $0 \leq r \lesssim 0.1$, the accuracy of the first model is limited (black dashed line), but by supplementing the $\Gamma_{Z}^{\mathrm{(loss)}}$ with the second-order corrections $\sim \kappa_{-}^2/\kappa_2$ (blue dotted line), see Eq.~\eqref{eq:bit_flip_rate_ordinary_cat}, the theory accurately captures the interplay of the competing mechanisms introducing bit-flip errors.

\section{Theoretical phase-flip rates for stabilized squeezed cats \label{app:phase_flip_rates}}

The main feature of the squeezed cat stabilization scheme 
presented in~\cite{xu_autonomous_2023}
is that the phase-flip rate can be reduced with an increasing value of the squeezing parameter $r$. We can briefly sketch their derivation. 
First of all, and for the sake of convenience, we are going to work in a squeezed frame. 
By means of a squeezed transformation, we remove the squeezing from the cat, but the loss operator becomes the squeezed annihilation operator $\hat{a} \to \hat{S}^\dagger(r) \hat{a} \hat{S}(r)$. 
In the squeezed frame, the combined stabilization and single-photon loss dynamics is described by the master equation
\begin{align}\label{app:squeezed-frame}
	\dv{}{t} \hat{\rho} = \kappa_2 \mathcal{D}[\hat{Z}_L (\hat{a}^2 - \beta^2)]\hat{\rho} + \kappa_{-} \mathcal{D}[\hat{S}^\dagger(r) \hat{a} \hat{S}(r)] \hat{\rho}.
\end{align}
In the squeezed frame, we can use the standard shifted Fock basis (SFB)~\cite{chamberland_building_2022} to describe the cat basis. 
Here the Hilbert space is split into a 2-dimensional logical space (the cat qubit manifold) and an auxiliary bosonic mode or gauge mode
with annihilation operator $\tilde{a}$.
In this basis and in the limit of large $\beta$, the bosonic annihilation operator can be approximated by 
\begin{align}
	\hat{a} \to \hat{Z}_L \otimes (\tilde{a} + \beta).
\end{align}
Notice from Eq.~\eqref{app:squeezed-frame} that the SFB in the squeezed frame is defined by the effective displacement $\beta$.
The SFB representation of the stabilization operator is
\begin{align}
	\hat{Z}_L (\hat{a}^2 - \beta^2) \to \hat{Z}_L \otimes \tilde{a} .
\end{align}
Similarly, the SFB representation of the squeezed annihilation operator is
\begin{align}
	\hat{S}^\dagger(r) \hat{a} \hat{S}(r) &= \hat{a} \cosh(r) - \hat{a}^\dagger \sinh(r) \nonumber\\
	&\to \hat{Z}_L \otimes \left[ \tilde{a} \cosh(r) - \tilde{a}^\dagger \sinh(r) + \beta \text{e}^{-r}  \right] \nonumber\\
    &= \hat{Z}_L \otimes \left[ \tilde{a} \cosh(r) - \tilde{a}^\dagger \sinh(r) + \alpha  \right] . \label{eq:squeezed-loss-sfb}
\end{align} 
From the expression above, only the last term $\alpha \hat{Z}_L \otimes \tilde{1}$ leads to uncorrectable phase-flip errors 
(logical phase-flips without any action on the gauge mode).
Note for example
that the term $\hat{Z}_L \otimes \tilde{a}^\dagger$, upon the action of the stabilization operator, reduces to the identity operator 
$(\hat{Z}_L \otimes \tilde{a})(\hat{Z}_L \otimes \tilde{a}^\dagger ) = \hat{Z}^2_L \otimes \tilde{a} \tilde{a}^\dagger \approx \hat{1}_L \otimes \tilde{1}$. This simplified analysis assumes that the gauge mode is initialized in the vacuum state, i.e., $\hat{\rho}(0) = \hat{\rho}_L \otimes \ketbra{\tilde{0}}$.
Therefore, residual uncorrectable phase-flip errors are created via
\begin{align}
	\kappa_{-} \mathcal{D}[\hat{S}^\dagger(r) \hat{a} \hat{S}(r)] \hat{\rho} &\to \kappa_{-} \mathcal{D}[ \alpha  \hat{Z}_L ] \hat{\rho}_L \\
	&= \kappa_{-} \alpha^2 \mathcal{D}[\hat{Z}_L] \hat{\rho}_L .
\end{align}
Using the approximation for the mean number of photons in the $\alpha \gg 1$ limit, $\alpha^2$ reduces to $\bar{n} - \sinh^2(r)$. Therefore, the phase-flip rate is given by 
\begin{align}
\label{eq:phase_flip_rate}
\Gamma_X = \kappa_{-} [\bar{n} - \sinh^2(r)]
\end{align}
Similarly, in the presence of dephasing noise with rate $\kappa_\phi$, 
i.e., $\kappa_\phi \mathcal{D}[\hat{S}^\dagger(r) \hat{a}^\dagger \hat{a} \hat{S}(r)]\hat{\rho}$,
the only terms giving rise to uncorrectable phase-flip errors (upon the action of the stabilization operation) in the SFB are those proportional to $\tilde{a}^\dagger$, that is, $\hat{1}_L \otimes \alpha [\cosh(r) - \sinh(r)] \tilde{a}^\dagger = \alpha {\rm e}^{-r} (\hat{1}_L \otimes\tilde{a}^\dagger)$. Therefore, uncorrectable phase-flip errors originating from a dephasing noise channel occur with a rate $\kappa_\phi {\rm e}^{-2r} [\bar{n} - \sinh^2(r)]$.

What is the difference with the model in Ref.~\cite{hillmann_quantum_2023}? Whereas the SFB of the squeezed annihilation operator is the same there is a small subtlety. The SFB representation of the stabilization operator in this reference is
\begin{align}
	 \hat{a}^2 - \beta^2 \to \hat{1}_L \otimes \tilde{a} .
\end{align}
This means that the term proportional to $\hat{Z}_L \otimes \tilde{a}^\dagger$ in \eqref{eq:squeezed-loss-sfb} which we neglected before as it did not lead to phase-flip errors plays a different role. Upon the subsequent action of the stabilization operator, the gauge mode is cooled down to the ground state with a residual phase-flip error $(\hat{1}_L \otimes \tilde{a})(\hat{Z}_L \otimes \tilde{a}^\dagger) = \hat{Z}_L \otimes \tilde{a} \tilde{a}^\dagger$. Thus, we need to consider the combined action of the terms
$\kappa_{-} \mathcal{D}[ \alpha \hat{Z}_L ]\hat{\rho}_L$, and $\kappa_{-} \mathcal{D}[ (-\sinh(r)) \hat{Z}_L ]\hat{\rho}_L$, which gives $\kappa_{-} \left( \alpha^2  + \sinh^2(r) \right) \mathcal{D}[\hat{Z}_L] \hat{\rho}_L$. 
Once again, in the $\alpha \gg 1$ limit, the above reduces to $\kappa_{-} \bar{n} \mathcal{D}[\hat{Z}_L] \hat{\rho}_L$.
Therefore, the phase-flip rate for this model is proportional to the mean number of photons.

\section{Stroboscopic reservoir engineering derivation \label{app:stroboscopic}}

Here we will derive a general framework for the stroboscopic reservoir engineering considering that the squeezed cat code is stabilized with the operators 
$\hat{S}_\Delta^{(2-\mu)}$, with $\mu = 0, 1$. 
Our results in main text correspond to the case $\mu=1$. Nevertheless, we also consider the case $\mu=0$ in order to compare our results to those in Ref.~\cite{shitara_exploiting_2025}.

Our starting point is the modular dissipator
\begin{align}
    \hat{d}^\mu_{\Delta} &= \frac{1}{\sqrt{2}} \left( \frac{\hat{q}_{[m_\mu]}}{\Delta} + i \Delta \hat{p}  \right) \pm \mu\frac{\alpha}{\Delta},
\end{align}
with $m_\mu = (1+\mu)\sqrt{2} \alpha$ the modularity that results from the periodicity of the stabilizer operators in the $q$ quadrature. 
As explained in the main text, our goal is to approximate the unitary evolution led by the 
interaction Hamiltonian 
\begin{align}
    \hat{H}_{\text{int}}(t) &= \sqrt{\Gamma} \left[ \hat{d}^\mu_{\Delta} \hat{w}^{\dagger}(t) + \hat{d}_{\Delta}^{\mu \dagger} \hat{w}(t) \right]  
\end{align}
between the bosonic mode in which we want to stabilize the logical qubit and a bath with operators satisfying the commutation relation $[\hat{w}(t), \hat{w}^\dagger(t')]= \delta(t-t')$.
In the above expression $\Gamma$ is the coupling strength which we assume to be constant (Markov approximation).

Now we will consider the discretized unitary evolution $\hat{U}(T) = \prod_{n=0}^N \hat{U}_n$, where $T = N \delta t$, with $T$ the total evolution time, $\delta t$ the small time step, and $N$ the maximum number of time steps. 
In this discretized model, the continuous bath operators are replaced by $\hat{w}(t) \to \hat{w}_n / \sqrt{\delta t}$, where $[\hat{w}_n, \hat{w}^\dagger_{n'}]=\delta_{nn'}$ and $t = n \delta t$. 
This leads to
\begin{align}
    \hat{U}_n &= \exp \left\{-i \sqrt{\Gamma \delta t} \left[ \hat{d}^\mu_\Delta \hat{w}^\dagger_n + \hat{d}^{\mu \dagger}_\Delta \hat{w}_n \right] \right\} \\
    &= \exp \left\{ -i \sqrt{\Gamma \delta t} \left[ \frac{1}{\Delta} \left( \hat{q}_{[m_\mu]} \pm \sqrt{2} \mu \alpha \right) \hat{q}_{n,w} \right. \right. \nonumber\\
    &+ \left. \left. \Delta \hat{p} \, \hat{p}_{n,w} \right] \right\} , 
\end{align}
where $\hat{q}_{n,w} = (\hat{w}^\dagger_n + \hat{w}_n)/\sqrt{2}$ and $\hat{p}_{n,w} = i(\hat{w}^\dagger_n - \hat{w}_n)/\sqrt{2}$ are the discrete bath mode position and momentum (respectively) quadrature operators.
We ``qubitize" $\hat{U}_n$ via the mapping $\hat{q}_{n,w} \to \hat{\sigma}_x / \sqrt{2}$, and $\hat{p}_{n,w} \to \hat{\sigma}_y /\sqrt{2}$, so that the commutation relation $[\hat{\sigma}_x / \sqrt{2}, \hat{\sigma}_y / \sqrt{2}] = i \hat{\sigma}_z $ approximates the canonical commutation relation in the limit $\langle \hat{\sigma}_z \rangle \approx 1$.
In the qubit model, the qubit plays the role of the $\hat{w}_n$ operators. 
Therefore, at each time step the qubit is initialized in its ground state and is reset back to it after the interaction with the bosonic mode takes place.
As the same qubit is being recycled, from now on we will abandon the $n$ index notation.

Thus, we are left with the single time-step unitary evolution 
\begin{align}\label{app:full-unitary}
    \hat{U} = \exp \left\{-i \sqrt{\frac{\Gamma \delta t}{2}} \left[ \frac{1}{\Delta} \left( \hat{q}_{[m_\mu ]} \pm \sqrt{2} \mu \alpha \right)\hat{\sigma}_x   + \Delta \hat{p}\hat{\sigma}_y \right] \right\} .
\end{align}
Compared to the unitary operator which gives rise to the (standard) sharpen-trim (ST), small-Big-small (sBs) and Big-small-Big (BsB) sequences, for $\mu = 1$ we have an additional term $ \propto \pm \sqrt{2} \alpha /\Delta $ which is a consequence of the minus sign in the stabilizer operator $\hat{S}_\Delta$~\cite{royer_stabilization_2020, shitara_exploiting_2025}. This offset will be realized via a qubit rotation.

In what follows, we will derive first- and second-order Suzuki-Trotter approximations to the unitary evolution \eqref{app:full-unitary}.

\subsection{First-order Suzuki-Trotter approximation - Sharpen and Trim sequences}\label{app:first-orderST}

First-order approximations can be obtained using the relation $\exponential(\hat{A}) \exponential(\hat{B}) \approx \exponential(\hat{A} + \hat{B} )$. 
This leads to the sharpen (S) and trim (T) gates
\begin{align}
    \hat{U}_{\text{S}} &= \exp \left(-i \sqrt{\frac{\Gamma \delta t}{2}}   \Delta \hat{p}\hat{\sigma}_y \right)  \nonumber\\
    &\times
    \exp \left[ -i \sqrt{\frac{\Gamma \delta t}{2}}  \frac{1}{\Delta} \left( 
    \hat{q}_{[ m_\mu ]} \pm \sqrt{2} \mu \alpha \right)\hat{\sigma}_x   \right] \\
    \hat{U}_{\text{T}} &= \exp \left[ -i \sqrt{\frac{\Gamma \delta t}{2}}  \frac{1}{\Delta} \left( \hat{q}_{[ m_\mu ]} \pm \sqrt{2} \mu \alpha \right)\hat{\sigma}_x  \right] \nonumber\\
    &\times
    \exp \left(-i \sqrt{\frac{\Gamma \delta t}{2}}   \Delta \hat{p}\hat{\sigma}_y \right) .
\end{align}
The final step consists in dealing with the modular position operator defined as
$\hat{q}_{[m_\mu]} = \hat{q} \mod m_\mu$. This definition implies that 
the position quadrature only takes values within an interval of length $m_\mu$ and therefore, shifts by multiples of $m_\mu$ can be ignored.
Consequently, we can perform the replacement $\hat{q}_{[m_\mu]} \to \hat{q}$ if we guarantee that the above unitaries are invariant under translations along the $q$ quadrature of amplitude equal to an integer multiple of $m_\mu$.
This condition results in 
\begin{align}\label{app:mod-condition-ST}
    \sqrt{\frac{\Gamma \delta t}{2}} = \frac{\Delta \pi}{m_\mu} , 
\end{align}
which fixes the value of $\sqrt{\Gamma \delta t}$. 
{
In other words, imposing translational invariance of the approximate unitary fixes the cooling rate of the stabilization sequence. This cooling rate is inversely proportional to $\alpha$ and decreases exponentially as squeezing increases.
Finally, the S and T unitaries reduce to
\begin{align}
    \hat{U}_{\text{S}} &= \exp \left( -i \frac{\epsilon_\mu}{2} \hat{p} \hat{\sigma}_y  \right) 
    \exp \left( -i \frac{\ell_\mu}{2} \hat{q} \hat{\sigma}_x \right) 
    \exp \left( - i \frac{\mu\pi}{2} \hat{\sigma}_x \right) ,\\
    \hat{U}_{\text{T}} &= \exp \left( -i \frac{\ell_\mu}{2} \hat{q} \hat{\sigma}_x \right) 
    \exp \left( - i \frac{\mu\pi}{2} \hat{\sigma}_x \right) 
    \exp \left( -i \frac{\epsilon_\mu}{2} \hat{p} \hat{\sigma}_y  \right)  ,
\end{align}
where we have defined
\begin{align}
    \ell_\mu &= \frac{2\pi}{m_\mu}, \\
    \epsilon_\mu &= \frac{2 \pi \Delta^2}{m_\mu}.
\end{align}
The larger modularity resulting from stabilizing the squeezed cat code with $\hat{S}_\Delta$ ($\mu=1$) leads to shorter conditional displacements (by a factor of 2) compared to the sequences obtained when using $\hat{S}_\Delta^2$ ($\mu=0$) as a stabilizer.

As already pointed out in Ref.~\cite{royer_stabilization_2020}, 
due to the presence of qubit operators, the above gate sequences are also invariant under displacements along the $p$ quadrature
$\text{e}^{i (2\pi/\epsilon_\mu) \hat{q}} \hat{U}_{\text{S(T)}} \text{e}^{-i (2\pi/\epsilon_\mu) \hat{q}} = -\hat{U}_{\text{S(T)}}$, with 
$2\pi/\epsilon_\mu = m_\mu / \Delta^2$ the so-called superlattice constant or the periodicity along the momentum quadrature. 

In circuit QED, the dispersive interaction between a bosonic mode and an auxiliary qubit allows to implement
displacements conditioned on the Pauli $Z$ eigenstates of an auxiliary qubit. This conditional displacement gate is defined as
\begin{align}
    \hat{\text{CD}}(\alpha) &= \exp \left[ \frac{1}{2\sqrt{2}} (\alpha \hat{a}^\dagger - \alpha^* \hat{a}) \hat{\sigma}_z \right] .
\end{align}
Using the relations $\hat{\sigma}_x = \hat{H} \hat{\sigma}_z \hat{H}$, and 
$\hat{\sigma}_y = \hat{R}_z(\pi/2) \hat{H} \hat{\sigma}_z \hat{H} \hat{R}^\dagger_z(\pi/2)$, where $\hat{H}$ is the Hadamard gate, it is possible to express the S and T sequences in terms of $Z$ conditional displacements as
\begin{align}
    \hat{U}_{\text{S}} &= \hat{\text{CD}}(\epsilon_\mu) \hat{R}^\dagger_x (\pi/2) \hat{\text{CD}}(-i\ell_\mu) \hat{R}_z (\mu \pi) ,\\
    \hat{U}_{\text{T}} &= \hat{\text{CD}}(-i\ell_\mu) \hat{R}_z (\mu \pi) \hat{R}_x (\pi/2)   \hat{\text{CD}}(\epsilon_\mu)  ,
\end{align}
where
\begin{align}
    \hat{R}_i (\theta) &= \exp (-i \theta \hat{\sigma}_i /2) ,
\end{align}
represents a qubit rotation by an angle $\theta$ around the angle defined by the Pauli matrix $\hat{\sigma}_i$.
Here we have discarded qubit rotations after the last conditional displacement as the qubit is to be reset. Both sequences start with the qubit in the Pauli $X$ +1 eigenstate.

\subsection{Second-order Suzuki-Trotter approximations}

Second-order approximations are obtained using the relation $\exponential(\hat{A}+\hat{B}) \approx \exponential(\hat{A}/2) \exponential(\hat{B}) \exponential(\hat{A}/2)$. 

\subsubsection{small-Big-small sequence}

The so-called small-Big-small (sBs) sequence follows by replacing $\hat{A} \propto \hat{p} \hat{\sigma}_y$ and $\hat{B} \propto \hat{q} \hat{\sigma}_x$. Thus, 
\begin{align}
    \hat{U}_{\text{sBs}} &= \exp \left(-i \sqrt{\frac{\Gamma \delta t}{2}} \frac{\Delta}{2} \hat{p}\hat{\sigma}_y \right)  \nonumber\\
    &\times
    \exp \left[ -i \sqrt{\frac{\Gamma \delta t}{2}}  \frac{1}{\Delta} \left( \hat{q}_{[ m_\mu ]} \pm \sqrt{2} \mu \alpha \right)\hat{\sigma}_x \right] \nonumber\\
    &\times
    \exp \left(-i \sqrt{\frac{\Gamma \delta t}{2}} \frac{\Delta}{2} \hat{p}\hat{\sigma}_y \right) ,
\end{align}
and after imposing translational invariance along the $q$ quadrature we are left with
\begin{align}
    \hat{U}_{\text{sBs}} &= 
    \exp \left( -i \frac{\epsilon_\mu}{4} \hat{p} \hat{\sigma}_y \right)
    \exp \left( -i \frac{\ell_\mu}{2} \hat{q} \hat{\sigma}_x \right) \nonumber\\
    &\times
    \exp \left( -i \frac{\mu \pi}{2} \hat{\sigma}_x \right) 
    \exp \left( -i \frac{\epsilon_\mu}{4} \hat{p} \hat{\sigma}_y \right)  .
\end{align}
Here, similarly to the first-order approximation case, translational invariance by integer multiples of $m_\mu$ along the $q$ quadrature yields $\sqrt{\Gamma \delta t/ 2} = \Delta \pi / m_\mu$ [cf. Eq.~\eqref{app:mod-condition-ST}]. 
That is, the cooling rates of both ST and sBs are the same.

The above sequence is also invariant under displacements along the $p$ quadrature.
In this case the superlattice constant $4\pi/\epsilon_\mu = 2 m_\mu / \Delta^2$ is twice as large as compared to the first-order approximation case due to the factor $1/2$ in the definition of the small gates ($\propto \hat{p} \hat{\sigma}_y$). 

The circuit-QED version of this sequence (once we have discarded the unnecessary qubit rotations) is
\begin{align}
    \hat{U}_{\text{sBs}} &= 
    \hat{\text{CD}}\left(\frac{\epsilon_\mu}{2}\right) 
    \hat{R}^\dagger_x \left(\frac{\pi}{2}\right) 
    \hat{\text{CD}}(-i\ell_\mu) 
    \hat{R}_z (\mu\pi) \nonumber\\
    &\times
    \hat{R}_x \left(\frac{\pi}{2}\right)
    \hat{\text{CD}}\left(\frac{\epsilon_\mu}{2}\right)  .
\end{align}
Once again, the qubit is initialized in the $\ket{+}$ state and is reset after the last conditional displacement.

\subsubsection{Big-small-Big sequence}

If instead we consider the replacements
$\hat{A} \propto \hat{q} \hat{\sigma}_x$ and
$\hat{B} \propto \hat{p} \hat{\sigma}_y$, 
one obtains the Big-small-Big (BsB) approximation
\begin{align}
    \hat{U}_{\text{BsB}} &= \exp \left[ -i \sqrt{\frac{\Gamma \delta t}{2}}  \frac{1}{2\Delta} \left( \hat{q}_{[ m_\mu ]} \pm \sqrt{2} \mu \alpha \right)\hat{\sigma}_x \right] \nonumber\\ 
    &\times
    \exp \left(-i \sqrt{\frac{\Gamma \delta t}{2}} \Delta \hat{p}\hat{\sigma}_y \right)  \nonumber\\
    &\times
    \exp \left[ -i \sqrt{\frac{\Gamma \delta t}{2}}  \frac{1}{2\Delta} \left( \hat{q}_{[ m_\mu ]} \pm \sqrt{2} \mu \alpha \right)\hat{\sigma}_x \right] .
\end{align}
After imposing translational invariance we are left with
\begin{align}
    \hat{U}_{\text{BsB}} &= 
    \exp \left( -i \frac{\ell_\mu}{2} \hat{q} \hat{\sigma}_x \right) 
    \exp \left( -i \frac{\mu \pi}{2} \hat{\sigma}_x \right) \nonumber\\
    &\times
    \exp \left( -i \epsilon_\mu \hat{p} \hat{\sigma}_y \right) \nonumber\\
    &\times
    \exp \left( -i \frac{\ell_\mu}{2} \hat{q} \hat{\sigma}_x \right) 
    \exp \left( -i \frac{\mu \pi}{2} \hat{\sigma}_x \right) 
      .
\end{align}
This time translational invariance along the $q$ quadrature results in 
\begin{align}
    \sqrt{\frac{\Gamma \delta t}{2}} &= \frac{2 \Delta \pi}{m_\mu} ,
\end{align}
while the superlattice constant becomes $m_\mu / 2\Delta^2$.
This means that while BsB has the smallest superlattice constant among the three studied sequences, it also has the largest cooling rate (twice that of ST and sBs).
In terms of circuit-QED conditional displacements, the above sequence reads 
\begin{align}
    \hat{U}_{\text{BsB}} &= 
    \hat{\text{CD}}(-i\ell_\mu)
    \hat{R}_z (\mu \pi)
    \hat{R}_x \left(\frac{\pi}{2}\right)
    \hat{\text{CD}}(2 \epsilon_\mu) 
    \hat{R}^\dagger_x \left(\frac{\pi}{2}\right) \nonumber\\
    &\times
    \hat{\text{CD}}(-i\ell_\mu) 
    \hat{R}_z (\mu \pi) .
\end{align}

\subsection{Numerical performance \label{app:numerical_performance}}
Different approaches to implementing the unitary Eq.~\eqref{app:full-unitary} yield different approximations and, consequently, different bit- and phase-flip error rates.
This stems from various factors.
First, the ST protocol is only a first-order approximation of Eq.~\eqref{app:full-unitary}, whereas sBs and BsB are second-order Trotter approximations. %
Second, the protocols also have different cooling rates and superlattice constants.
Royer \emph{et al.}~\cite{royer_stabilization_2020} found that the optimal sequence depends on the operating regime and the type of oscillator decoherence.
For example, for single-photon loss rates $\kappa \delta t > 10^{-3}$ the BsB sequence performs better than the sBs sequence~\cite[Fig. 3a]{royer_stabilization_2020}, likely due to the higher cooling rate that is more efficient in removing excitation errors.
For $\kappa \delta t < 10^{-3}$, the sBs sequence outperforms the BsB sequence, due to the poorer confinement properties of BsB caused by the smaller superlattice constant.

For stabilized squeezed cat qubits, we do find that this intuition partially carries over. 
On the one hand, if the noise source is perturbative, the sBs sequence significantly outperforms BsB and ST.
The reduced confinement properties of BsB and ST lead to a significantly increased bit-flip error rate $\Gamma_{Z}$ compared to the sBs sequence, see \sfigref{fig:sequence_comparison_combined}{a}.
Note that the slopes of $ \Gamma_{Z} $ also differ significantly between the sBs sequence on the one side, and the ST and BsB sequences that have identical slopes.
We note that in terms of the phase-flip rate, the sequences show a reversed order in performance, see \figref{fig:sequence_comparison_combined}.
However, relative improvements $\Gamma_{X}^{\mathrm{sBs}} / \Gamma_{X}^{\mathrm{BsB}}$ are on the order of 1.

On the other hand, the increased cooling rate of the BsB sequence gives an advantage in the regime of strong noise.
By strong noise, we mean the regime in which the bit-flip rate $\Gamma_{Z}$ becomes a nonlinear function of the noise rate $\kappa \delta t$, see also \figref{fig:sBs_loss_sweep}.
We compare the performance of the different stabilization sequences for $\kappa \delta t = 10^{-2}$ in \figref{fig:app_sBs_BsB_ST_high_noise_loss}.
There, the difference in the cooling rate of the BsB and ST stabilization sequences is already evident for small squeezing ($r=0.2$ and $r=0.4$).
While the sBs sequence outperforms the BsB and ST sequences significantly when the noise remains perturbative, we find that BsB can outperform sBs in the strong noise regime due to its increased cooling rate.
Indeed, we find that in this regime BsB leads to the lowest bit-flip rates.

Lastly, we note that we do not find the above-mentioned feature to be universal across noise sources. 
For example, in the case of strong photon-number dephasing noise (not shown), the BsB sequence can outperform the sBs sequence, but the results are not as clean as in \figref{fig:app_sBs_BsB_ST_high_noise_loss}.

\begin{figure}
    \centering
    \includegraphics{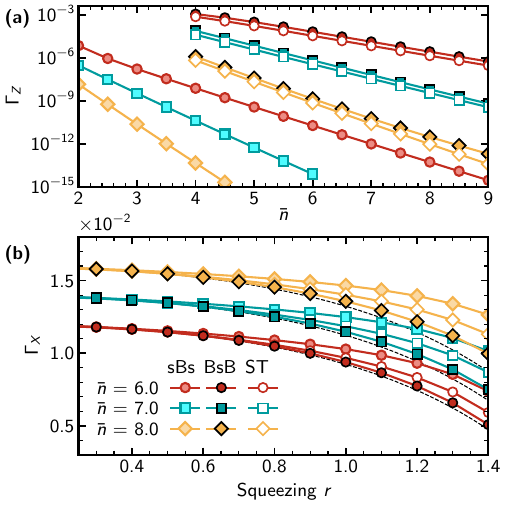}
    \caption{Bit- and phase-flip error rates of different stabilization sequences.
    Simulations are performed under the assumptions of perfect and instantaneous gates interleaved by evolution under the single-photon loss channel with  $\kappa \delta t= 10^{-3}$.
    The red, teal, and yellow lines in \textbf{(a)} correspond to squeezing levels $r=0.0$, $r=0.2$, $r=0.4$, respectively.
    The marker style determines the stabilization sequence, see (b).
    The black dashed lines in \textbf{(b)} correspond to the optimally achievable phase-flip rates for a given average photon number $\bar{n}$, see \secref{app:phase_flip_rates}.
}
    \label{fig:sequence_comparison_combined}
\end{figure}

\begin{figure}[!ht]
    \centering
    \includegraphics{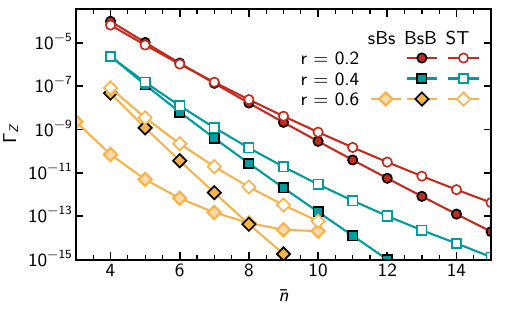}
    \caption{ Performance of stabilization sequences in the regime of strong single-photon losses with $\kappa \delta t = 10^{-2}$.
    In this regime, the noise becomes non-perturbative, and protocols with a reduced cooling rate exhibit a non-exponential reduction in the bit-flip rate.
    }
    \label{fig:app_sBs_BsB_ST_high_noise_loss}
\end{figure}

\section{Phase-space portraits of ST and BsB}\label{app:phase-portraits}

For the sake of completeness, in \sfigref{fig:phase_portrait_BsB_ST}{a} and \sfigref{fig:phase_portrait_BsB_ST}{b}, we show the phase-space portrait of the ST sequence, and in 
\sfigref{fig:phase_portrait_BsB_ST}{c} and \sfigref{fig:phase_portrait_BsB_ST}{d}, the phase-space portrait of the BsB sequence, both of them derived from the $\hat{S}_\Delta$ stabilizer operator. 
As detailed above, the superlattice constant (periodicity along the $p$ quadrature) of ST is $m_1/ \Delta^2$, and that of BsB is $m_1/ 2\Delta^2$, with $m_1 = 2\sqrt{2} \alpha$ the periodicity along the $q$ quadrature.
For the same stabilization sequence, the superlattice constant can be increased by means of the squeezing parameter $\Delta = \exp(-r)$. 
For ST, we compare the zero-squeezing and non-zero-squeezing cases in \sfigref{fig:phase_portrait_BsB_ST}{a} and \sfigref{fig:phase_portrait_BsB_ST}{b}, respectively. Similarly, for BsB, we compare the zero-squeezing and non-zero-squeezing cases in \sfigref{fig:phase_portrait_BsB_ST}{c} and \sfigref{fig:phase_portrait_BsB_ST}{d}, respectively.

\begin{figure}[!ht]
    \centering
    \includegraphics{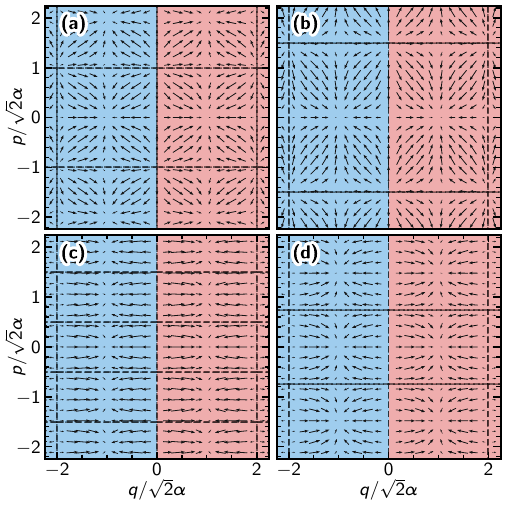}
    \caption{
    Phase-space portraits of ST and BsB stabilization protocols.
    In all panels $\alpha = \sqrt{3}$, with $r = 0$ in (a) and (c), and $r = 0.2$ in (b) and (d).
    The top panels (a) and (b) correspond to the ST protocol and the bottom panels (c) and (d) correspond to the BsB protocol.
    Dashed lines indicate lattice and superlattice periodicity.
    The color of each cell encodes whether starting in this cell yields dynamics that are projected into $\ket{0}$ (red) or $\ket{1}$ (blue) upon measurement.
    }
    \label{fig:phase_portrait_BsB_ST}
\end{figure}

\section{Observables of the encoded state \label{app:observables}}
In order to estimate the effective phase- and bit-flip rates of the stroboscopically stabilized logical states, 
we require observables that reveal both the parity of the states, and whether their support lies in the positive or negative half-plane of phase space.

Concretely, we relate the expectation value of Pauli $X$ ($\sigma_X$)
to the bosonic parity operator which we denote here as $\hat{J}_{X}$ and is given by
\begin{align}
    \hat{J}_X = \hat{J}_{++} - \hat{J}_{--},
\end{align}
with 
\begin{align}
    \hat{J}_{++} &= \sum_{n=0}^{\infty} \ketbra{2n},\\
    \hat{J}_{--} &= \sum_{n=0}^{\infty} \ketbra{2n + 1}, 
\end{align}
and therefore, 
$\expval{\sigma_X} = \Trace[\hat{J}_{X} \hat{\rho}]$.
On the other hand, 
for Pauli $Z$ ($\sigma_Z$)
we use an adaption of the observable introduced in Ref.~\cite{mirrahimi_dynamically_2014} which is a good approximation of $\mathop{\mathrm{sign}}(\hat{a} + \hat{a}^{\dagger})$ and is defined as 
\begin{align}
    \hat{J}_Z = \hat{J}_{+-} + \hat{J}_{+-}^{\dagger},
\end{align}
with 
\begin{align}
    \label{eq:app_j_pm}
    \hat{J}_{+-}=\sqrt{\frac{2 \alpha^2}{\sinh \left(2 \alpha^2\right)}} \sum_{q=-\infty}^{\infty} \frac{(-1)^q}{2 q+1} I_q(\alpha^2) \hat{J}_{+-}^{(q)},
\end{align}
where $\alpha \in \mathbb{R}$, $I_q(x)$ is the modified Bessel function of the first kind and $\hat{J}^{(q)}_{+-}$ is further defined as
\begin{align}
\hat{J}_{+-}^{(q)}=\left\{\begin{array}{ll}
\dfrac{\left(\hat{a}^{\dagger} \hat{a}-1\right) ! !}{\left(\hat{a}^{\dagger} \hat{a}+2 q\right) ! !} \hat{J}_{++} \hat{a}^{2 q+1} \quad q \geq 0 \\
\hat{J}_{++} \hat{a}^{\dagger(2|q|-1)}   \dfrac{\left(\hat{a}^{\dagger} \hat{a}\right) ! !}{\left(\hat{a}^{\dagger} \hat{a}+2|q|-1\right) ! !} & q<0
\end{array}\right.,
\end{align}
where $n!! = \prod_{k=0}^{\lfloor (n-1)/2 \rfloor} (n-2k)$ denotes the double factorial.
Thus, $\expval{\sigma_Z} = \Trace[\hat{J}_{Z} \hat{\rho}]$.

While $\hat{J}_{Z}$ correctly captures the decay rate of the SCQ basis states, 
it is desired to correctly normalize this operator so that the basis states become its $\pm 1$ eigenstates. 
To this end, we compute instead
$\hat{S}(r) \hat{J}_Z \hat{S}^{\dagger}(r)$ while additionally replacing $\alpha^2$ with $\beta^2 = \alpha^2 e^{2 r}$ in $\hat{J}_{+-}$ [Eq.~\eqref{eq:app_j_pm}].

\subsection{Estimating error rates}
We numerically calculate the effective bit- and phase-flip rates by fitting the decaying logical observables,
\begin{align}
   \Trace[\hat{J}_Z \hat{\rho}(t)] &\sim \mathrm{e}^{-\Gamma_{Z} t}, \\
    \Trace[\hat{J}_X \hat{\rho}(t)] &\sim \mathrm{e}^{-\Gamma_{X} t}.
\end{align}
To this end, we initialize the system in an ideal state 
$\ket{\alpha, r}$ or $\ket{0}_\mathcal{C}$ (respectively)
before time evolving it for a total time $T = n \delta t$, with $n$ the total number of correction cycles.
For the case of ideal gates, $\delta t$ is the idle time in which the state is subject to the loss channel upon which the instantaneous ideal gates are applied. 
For the case of conditional displacement gates implemented via a Hamiltonian, $\delta t$ corresponds to the total duration of the stabilization sequence in which the system evolves under the combined action of the unitary control and the loss channel. 

The simulations are performed using the \texttt{qutip} package~\cite{lambert_qutip_2026}, then the data is analyzed and visualized utilizing \textsc{python} libraries~\cite{hunter_matplotlib_2007, harris_array_2020, virtanen_scipy_2020}.

\section{Oscillator errors - dephasing}\label{app:oscillator-dephasing}

Suppose we want to implement a conditional displacement gate of amplitude $g t_{\rm CD}$ in the presence of a dephasing noise channel 
\begin{align}
    \dv{}{t} \hat{\rho} = - i \comm{\hat{H}_{\rm CD}}{\hat{\rho}} + \kappa_\phi \mathcal{D}[\hat{a}^\dagger \hat{a}] \hat{\rho},
\end{align}
with $\hat{H}_{\rm CD}$ given in Eq.~\eqref{eq:cd_hamiltonian}, and $\kappa_\phi$ the dephasing noise rate.
In the regime, $\kappa_\phi t_{\rm CD} \ll 1$, we treat photon dephasing perturbatively by considering the effect of
a single $\hat{a}^\dagger \hat{a}$ error occurring mid-gate at time $\tau$ 
\begin{align}
    \hat{D}(g \bar{\tau}) \hat{a}^\dagger \hat{a} \hat{D}(g \tau) &= \left[ \hat{a}^\dagger \hat{a} - \lambda(\hat{a} + \hat{a}^\dagger) + \lambda^2 \right] \hat{D}(g t_{\rm CD}),
\end{align}
with $\lambda = g t_{\rm CD} (1 - \tau/t_{\rm CD})$ and $\bar{\tau} = t_{\rm CD} - \tau$. 
Similarly to the case of single-photon losses studied in the main text, the first term on the right-hand side corresponds to the action of $\hat{a}^\dagger \hat{a}$ after an instantaneous gate, while the other two terms represent residual displacements. 
The relative amplitudes between the first and the other two terms are $\sqrt{\bar{n}}/\lambda$ and $\sqrt{\bar{n}}/\lambda^2$, respectively. 
In the parameter regimes studied in this work, these contributions are not significant, and therefore, we do not expect major deviations from the results presented in the main text, which consider instantaneous gates.

%

\end{document}